\documentclass[a4paper,fleqn,usenatbib]{mnras}

\usepackage{mathptmx}

\usepackage[T1]{fontenc}
\usepackage{ae,aecompl}

\usepackage{graphicx}   
\usepackage{amsmath}    
\usepackage{amssymb}    

\usepackage{booktabs}

\usepackage[T1]{fontenc}
\usepackage{aecompl}

\usepackage{multirow}

               {           \\\bottomrule%
                     \end{tabular}%
                   \end{small}
               \end{center} 
               }

\usepackage{dblfloatfix}

\usepackage[usenames]{color}

\usepackage[normalem]{ulem}

\newcommand{\be}{\begin{equation}}
\newcommand{\ee}{\end{equation}}
\newcommand{\bi}{\begin{itemize}}
\newcommand{\ei}{\end{itemize}}

\usepackage{booktabs, multirow, array}

\def\bmp#1{\begin{minipage}[!t]{#1\textwidth}}
\def\emp{\end{minipage}}

\usepackage{multicol}

\usepackage{afterpage}  
\usepackage{placeins}

\def\aline{
  \noindent\makebox[\textwidth]{\rule{\textwidth}{0.4pt}}
}

\usepackage{float} 

\title[Disformal quintessence]{Non-linear Phenomenology of Disformally Coupled Quintessence}

\author[Llinares, Hagala and Mota]{
Claudio Llinares,$^{1,2}$\thanks{E-mail: claudio.llinares@port.ac.uk}
Robert Hagala$^{3}$, 
and David F.~Mota$^{3}$
\\
$^{1}$Institute of Cosmology and Gravitation, University of Portsmouth, Dennis Sciama Building, Portsmouth PO1 3FX, United Kingdom \\
$^{2}$Institute for Computational Cosmology, Department of Physics, Durham University, Durham DH1 3LE, U.K. \\
$^{3}$Institute of Theoretical Astrophysics, University of Oslo, PO Box 1029 Blindern, 0315 Oslo, Norway
}

\date{Accepted XXX. Received YYY; in original form ZZZ}

\pubyear{2019}

\begin{document}
\label{firstpage}
\pagerange{\pageref{firstpage}--\pageref{lastpage}}
\maketitle

\begin{abstract}
The Quintessence model is one of the simplest and better known alternatives to Einstein's theory for gravity.  The properties of the solutions have been studied in great detail in the background, linear and non-linear contexts in cosmology.  Here we discuss new phenomenology that is induced by adding disformal terms to the interactions.  Among other results, we show analytically and using cosmological simulations ran with the code \texttt{Isis} that the model posses a mechanism through which is it possible to obtain repulsive fifth forces, which are opposite to gravity.  Although the equations are very complex, we also find that most of the new phenomenology can be explained by studying background quantities.  We used our simulation data to test approximate relations that exist between the metric and scalar field perturbations as well as between the fifth force and gravity.  Excellent agreement was found between exact and approximated solutions, which opens the way for running disformal gravity cosmological simulations using simply a Newtonian solver.  These results could not only help us to find new ways of testing gravity, but also provide new motivations for building alternative models.  
\end{abstract}




\section{Introduction \label{sec:introduction}}

Observations of distant supernovae, quasars, and of the Cosmic Microwave Background are consistent with a universe with late time accelerated expansion (\cite{FirstAccel,HubbleSN,HubbleQuasars,Planck2015}).  Although this effect can be mimicked by introducing a cosmological constant to Einstein's equations, the true nature of the expansion is still unknown. Moreover, the apparent value of the cosmological constant does not correspond to the vacuum energy predicted by particle physics \citep[e.g.][]{CosmologicalConstantProblem}.  Among the several solutions to these inconsistencies, there is the idea of modifying Einstein's theory for gravity.  A comprehensive description of many of these theories and their cosmological implications can be found in reviews by \cite{ModGravCosm}, \cite{2018LRR....21....2A}, \cite{2006IJMPD..15.1753C}, \cite{2009RPPh...72i6901S}, \cite{2013PhR...530...87W}, \cite{2015PhR...568....1J}, \cite{2004PhRvD..69d4005L}, \cite{2016RPPh...79d6902K}, \cite{2017PhR...692....1N} or \cite{2018IJMPD..all}.

Many of these modified gravity (MG) theories can be interpreted as having two geometries for space-time.  One of these two geometries characterizes the curvature of space-time, while the other describes the impact that this curvature has on the dynamics of matter.  The simplest way of relating these two metrics is through a conformal transformation (i.e. one metric is obtained from the other with a rescaling).  This rescaling factor is equal in all dimensions and hence conserves shapes.  The next step in complexity consists in adding a dependence with the direction to this relation.  This can be done through so called disformal transformations, which in the case of scalar-tensor theories depend on the derivatives of a scalar field \citep{Bekenstein1993}.  These kind of transformations have been studied in several contexts in cosmology such as inflation \citep{DisformalInflation}, dark matter \citep{skordis,MimeticPurnendu}, dark energy \citep{DisformalQuintessenceKoivisto, Koivisto:2008,MifsudBackground,2015PhRvD..91b4036S}, screening of modified gravity \citep{DisformalScreening, 2015JCAP...10..051I}, non-linear structure formation \citep{DisformalSymmetron} and others \citep{barrow,brax,disformal1,disformal2,Achour,Mifsud2016,SaksteinD1, 2015PhRvD..92l3005S}.

In this work, we will study the impact that the addition of a disformal coupling has on the solutions of field equations and the non-linear matter distribution.  We give a concrete example which we obtained by perturbing with a disformal coupling the quintessence model, which is one of the best known extended models of gravity.  Contrary to other works in which the emphasis is put on parameter estimation, here we are interested in finding novel phenomenology associated to this coupling, independently of the validity of the model from an observational perspective.  Doing this is important because knowing what effects are associated to this particular coupling could help to construct models with similar phenomenology, but that are compatible with specific data sets.  This in turn may enable us to construct novel tests of gravity based on this new phenomenology.

Among other characteristics of the solutions of the Klein-Gordon equation for the quintessence field, we will discuss a very simple relation that the disformal coupling induces between this field and the gravitational potential, which can be translated into a similar relation between the fifth force associated to the scalar field and gravity.  We will first analyse this and other effects analytically.  We will confirm these estimations a posteriori in a realistic set up given by fully non-linear cosmological simulations.  These simulations track the scalar field by means of a non-linear hyperbolic solver which takes into account time derivatives in the background as well as in the perturbations and thus, provides the most accurate solution can that be obtained, without assuming specific symmetries or neglecting terms in the equations.

We present details of the model in section \ref{sec:model}.  In section \ref{section:analytical} we describe analytical properties of the evolution of the background and perturbed scalar field as well as of the fifth force that arises from it.  Section \ref{section:nbody} describes the cosmological simulation suit in which we base our non-linear analysis.  We present results from these simulations on the scalar field and matter distributions in Sections \ref{sec:simulation_results_field} and \ref{sec:matterresults} respectively.   We summarize our results and conclude in section \ref{sec:conclusions}.

\section{Disformally coupled quintessence} 
\label{sec:model}

The model that we consider in this work can be defined with the following action
\begin{equation}
S=\intop \!\left[\sqrt{-g}\left(\frac{R}{16\pi G}+X-V \left(\phi\right)\right) + \sqrt{-g} \mathcal{L}_{B} + \sqrt{-\tilde{g}} \tilde{\mathcal{L}}_{DM} \right]\mathrm{d}^{4}x,\label{eq:action}
\end{equation}
where $X$ is the kinetic energy density of the field, defined by
\begin{equation}
X\equiv -\frac{1}{2} \phi^{,a}\phi_{,a}
\end{equation}
and $\mathcal{L}_{B}$ and $\tilde{\mathcal{L}}_{DM}$ are the Lagrangians of the baryonic and dark matter fields respectively.  We assume that the coupling is non-universal and that these two fields are coupled to the Einstein and Jordan frames metrics $g_{ab}$ and $\tilde{g}_{ab}$ respectively, which are related through the following disformal transformation:
\begin{equation}
\tilde{g}_{ab}=g_{ab}+B\left( \phi \right) \phi_{,a}\phi_{,b}. \label{eq:coupling}
\end{equation}
The reason for adopting a non-universal coupling is that the model does not include a screening mechanism.  With this assumption, we ensure that only the dark matter component of the Universe will be affected by the modification to gravity and that Solar System constraints will be fulfilled.   Furthermore, the choice of a non-universal coupling also ensures that the model is compatible with recent constraints on the speed of gravity waves that were obtained through the detection of an optical counterpart of a black-hole collision \citep{GW_gamma}.  Note that a similar approach was followed by several authors already \citep{2013JCAP...11..022X, 2011PhRvD..83b4007L, 2015PhRvD..91h3537S, 2015Ap&SS.359...11B, MorriceVDB}.  Screening mechanisms may be added by making appropriate choices of the conformal part of the transformation, which we assume is equal to one.  An example of such a procedure was presented by \cite{DisformalSymmetron}, who studied the effects of a disformal coupling included on top of the symmetron model, which is defined by a conformal coupling $1+\phi^2$. 

We choose the following form for the disformal coupling $B$ and the potential $V$
\begin{align}
B\left( \phi \right) & = B_0 \exp \left( \beta \phi / M_\mathrm{pl} \right ), \label{eq:B} \\
V\left( \phi \right) & = V_0 \exp \left( - \nu \phi / M_\mathrm{pl} \right ), \label{eq:V}
\end{align}
which were already studied on several occasions by \cite{DisformalQuintessenceKoivisto}, \cite{DisformalScreening}, \cite{DBIgalileons}, \cite{MorriceVDB}, \cite{Mifsud2016}, \cite{2010JCAP...05..038Z} and \cite{2015PhRvD..91b4036S}. This choice was made not only because it provides simple equations from which several analytical properties can be studied, but also because a change of the initial value of the scalar field, $\phi \rightarrow \phi_\mathrm{init} + \bar \phi $, can be collected into a change of the parameters $B_0$ and $V_0$. Taking this into account releases us from treating the initial value of the field as a free parameter, allowing us to fix $\phi_\mathrm{init} = 0$ without loss of generality.

\begin{table*}
\centering
\begin{tabular}{lll}
  
    Type of transition & Without damping & With damping \\
  \hline
  Quintessence linear $\rightarrow$ Quintessence non-linear & 
  $T^\mathrm{nd}_a \equiv 2\sqrt{2}$ & 
  $T_a \equiv 5.7$ (numerical) \\
  Disformal linear $\rightarrow$ Quintessence linear & 
  $T^\mathrm{nd}_b \equiv \sqrt{6D}$ & 
  $T_b \equiv \sqrt{2D} $ \\
  Disformal linear $\rightarrow$ disformal non-linear & 
  $T^\mathrm{nd}_c \equiv 3^{7/12}\Gamma^{1/3}\left(3/4\right)\sqrt{2}D^{1/4}$ & 
  $T_c \equiv T^{\mathrm{nd}}_c$ 
\end{tabular}
\caption{Characteristic time scales that arise in the background solutions of the Klein-Gordon equation for Einstein-de Sitter cosmology.  Note that the non-linear regimes mentioned here are associated with the moment in which the background equation for the scalar field becomes non-linear (and not to the non-linear regime usually studied in cosmology).  The superscript nd makes reference to ``non-damped'' solutions.}
\label{tab:definitions}
\end{table*}


We will restrict our analysis to positive values of $B_0$, $V_0$, and $\nu$. The potential $V(\phi)$ is thus a decreasing function of $\phi$, which will result in a background field rolling down the potential towards infinity.  Negative values of $\nu$ would simply make the field roll towards negative values instead, and the analysis in this paper would be identical after the transformation $\phi \rightarrow -\phi$ and $\beta \rightarrow -\beta$.  A positive choice of $B_0$ is needed to ensure $B\left( \phi \right) > 0$. As mentioned by \cite{Bekenstein1993}, a negative coupling $B\left( \phi \right)$ breaks causality by allowing information in the scalar field to propagate faster than the speed of light.  We will look at three different general cases for $\beta$: negative, positive, and zero. These correspond to a disformal coupling $B(\phi)$ which is respectively decreasing with $\phi$, increasing with $\phi$, or constant.

Variation of the action \eqref{eq:action} with respect to the field $\phi$ yields the following equation of motion for the scalar field:
\begin{equation}
\ddot{\phi}=\frac{1}{\left(1+\gamma^2\rho\right)}  \left[ \frac{c^2}{a^2}\nabla^{2}\phi -3H\dot \phi - \frac{1}{2} B_{,\phi} \rho \dot{\phi}^{2} -V_{,\phi} \right], \label{eq:EOM}
\end{equation}
where
\be
\gamma^2 = \frac{B}{1 + B \phi^{,a}\phi_{,a}},
\ee
we assumed that matter is a pressureless perfect fluid and that the Einstein frame metric takes the following form:
\begin{equation}
\mathrm{d}s^{2}=-\left(1+2 \Psi \right)\mathrm{d}t^{2}+a^{2}\left(t\right)\left(1-2 \Psi\right)\left(\mathrm{d}x^{2}+\mathrm{d}y^{2}+\mathrm{d}z^{2}\right),
\label{metric}
\end{equation}
where the Newtonian frame scalar perturbation $\Psi$ is the usual Newtonian potential.  Here and throughout this paper, a dot corresponds to a partial derivative with respect to cosmic time $t$.  Note that the only differences that the equation of motion (\ref{eq:EOM}) has with respect to the usual quintessence model are:
\bi
\item A factor $\left(1+\gamma^2\rho\right)^{-1}$ which changes both the speed at which the scalar field evolves in the background and the speed of scalar waves.
\item The addition of a term $\frac{1}{2} B_{,\phi} \rho \dot{\phi}^{2}$ (i.e. an additional force acting on the field).
\ei
These two additional terms are not exclusive to the base model we choose (in this case the quintessence model), but are characteristic of the disformal coupling.  The aim of this paper is to understand the consequences that these terms have in both the solutions of the Klein-Gordon equation and the non-linear distribution of matter in the Universe.

\section{Analytical Properties of the Model \label{sec:analytical}}
\label{section:analytical}

This section describes analytical solutions of the Klein-Gordon equation for the scalar field.  We will study separately the time evolution of the scalar field in the background (for an Einstein-de Sitter universe) and its perturbations.  Furthermore, we will discuss properties of the fifth force that arise from it.

\subsection{Disformal field dynamics: background evolution in an Einstein-de Sitter universe} 
\label{background_evolution}

\begin{figure*}
  \includegraphics[width=\textwidth]{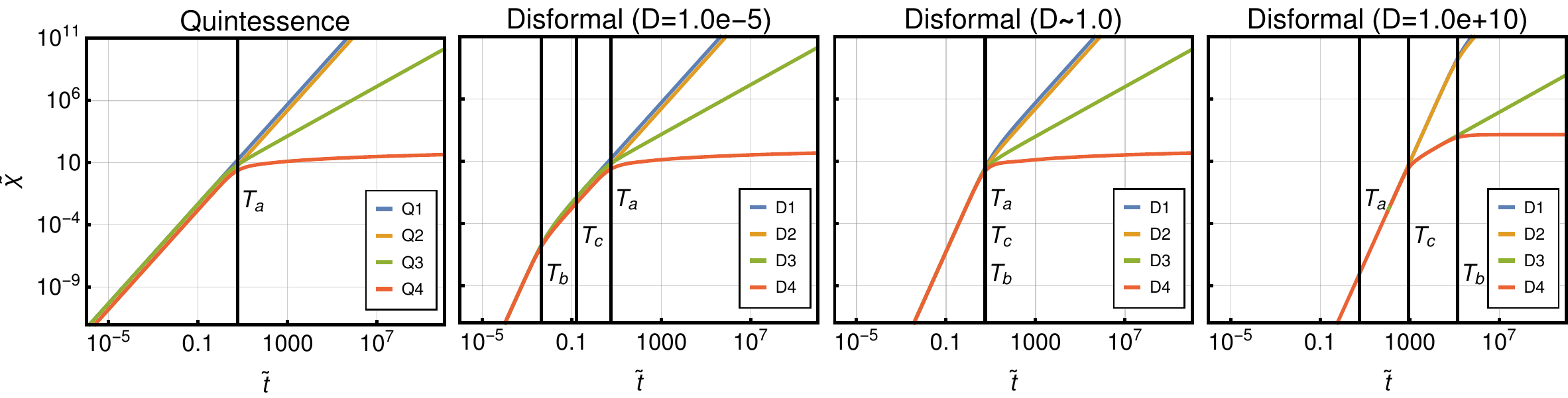}
  \caption{Background evolution of the field for quintessence model (left panel) and disformal model with constant coupling (for three different values of the free parameter $D$ and $F=0$).  Each panel contains four curves that correspond to the four models described in Tables \ref{tab:quintessence} and \ref{tab:disformal_beta_equal_zero} in Appendix \ref{appendix:table}.  The vertical lines correspond to the different time scales defined in Table \ref{tab:definitions} for the case that includes damping.   The numerical values of these time scales are very close to the values that do not include damping.}
  \label{fig:four_background_solutions} 
\end{figure*}

Before studying properties of the solutions, it is convenient to write the Klein-Gordon equation (\ref{eq:EOM}) with dimensionless variables.  In the limit $\gamma^2\rightarrow B$, the equation takes the following form:
\begin{multline}
\partial_{\tilde{t}}^2\tilde{\chi} = 
-2\frac{\tilde{t}}{\tilde{t}^2 + D\exp(F\tilde{\chi})}\partial_{\tilde{t}}\tilde{\chi} - 
\frac{1}{2}\frac{FD}{\tilde{t}^2+D\exp(F\tilde{\chi})} \left(\partial_{\tilde{t}}\tilde{\chi}\right)^2 +  \\
\frac{\tilde{t}^2}{\tilde{t}^2+D\exp(F\tilde{\chi})}\exp\left(-\tilde{\chi}\right), 
\label{eom_back}
\end{multline}
where we used the following dimensionless variables: 
\begin{align}
  \tilde{\chi} & \equiv \nu\frac{\phi}{M_\mathrm{P}},  \\
  \tilde{t} & \equiv \sqrt{v_0}\nu H_0 t
\end{align}
and the following dimensionless parameters:
\begin{align}
\label{defs_1}
b_0 & \equiv H_0^2 M_\mathrm{pl}^2 B_0, &
D & \equiv \frac{4}{3} b_0 v_0 \nu^2, \\
\label{defs_2}
v_0 & \equiv \frac{V_0 }{H_0^2 M_\mathrm{pl}^2},  &
F & \equiv \frac{\beta}{\nu}.
\end{align}
Cosmological energy scales are of the order of $H_0^2 M_\mathrm{pl}^2$, meaning that these rescalings will give cosmological consequences for model parameters $b_0$ and $v_0$ close to unity.  Since details of the background evolution of the metric are not expected to change the phenomenology provided by the disformal terms, we will study the simplest case: a flat universe with no cosmological constant.  Thus, we derived Eq.~\ref{eom_back} by assuming the following relations between time and expansion factor
\be
a(t) = \left(\frac{3 H_0 t}{2}\right)^{2/3}
\ee
and the following evolution of the background density
\be
\label{rho_background}
\rho(a) = \frac{\rho(a=1)}{a^3} = \frac{3H_0^2 M_P^2}{a^3}.
\ee
Note that this solution is not strictly valid in disformal gravity.  However, taking into account corrections associated with the disformal coupling will add a new layer of complexity which is beyond the scope of this paper.

These definitions show that the four original free parameters $V_0, \nu, B_0$ and $\beta$ are degenerate and that the shape of the background solutions depends only on two free parameters D and F.  The limit $(D,F)\rightarrow 0$ corresponds to the usual quintessence model which does not depend on any free parameter (all the information provided by the two original parameters $V_0$ and $\nu$ can be condensed in the rescaling of the time and the scalar field).  The limit $F\rightarrow 0$ is associated with a constant disformal coupling, where $\beta=0$.  We will study these two limits first separately and then the most general case with $\beta \ne 0$.  

\subsubsection{Quintessence}
\label{sec:quint}

The quintessence limit is defined by assuming $(D,F) \rightarrow 0$ and gives rise to the following Klein-Gordon equation for the evolution of the background field:
\be
\partial_{\tilde{t}}^2\tilde{\chi} = -\frac{2}{\tilde{t}}\partial_{\tilde{t}}\tilde{\chi} +  \exp\left(-\tilde{\chi}\right).
\label{eq:quintessence}
\ee
The solution is shaped by the presence of a damping term and of a non-linear regime which is triggered when the field is large enough to reach the non-linear part of the exponential function.  In order to understand consequences of two effects, we study four different cases that correspond to equations that are and are not linearized with respect to the field and with and without the addition of the damping term.  Table \ref{tab:quintessence} in Appendix \ref{appendix:table} summarizes properties of these four solutions.  The definition of the time scales that appear in these results is given in Table \ref{tab:definitions} in this section.  The left panel of Fig.~\ref{fig:four_background_solutions} shows the evolution of the field in these four special cases.

The complete solution of the quintessence equation of motion (Eq. \ref{eq:quintessence}; red curve in Fig.~\ref{fig:four_background_solutions}) has a characteristic time scale that divides the linear regime at early times from the non-linear regime at late times.  At early times, the damping term interacts with the force that induces the field to roll down the potential in such a way that only the normalization of the solution is changed with respect to the undamped solution.  This regime is characterized by a logarithmic slope ($\frac{d\log\tilde{\chi}}{d\log\tilde{t}} = \frac{\tilde{t}}{\tilde{\chi}}\partial_{\tilde{t}}\tilde{\chi}$) equal to two.  During the transition to the non-linear regime, the force that accelerates the field becomes negligible and thus, the evolution of the field is damped and approaches a solution with a logarithmic slope equal to zero.

\subsubsection{Disformal gravity with constant disformal coupling}
\label{disformal_background_constant_coupling}

We now study solutions of the Klein-Gordon equation in the limit $F\rightarrow 0$ and $D\neq 0$:
\be
\label{disformal_no_beta}
\partial_{\tilde{t}}^2\tilde{\chi} = -2\frac{\tilde{t}}{\tilde{t}^2+D}\partial_{\tilde{t}}\tilde{\chi} + \frac{\tilde{t}^2}{\tilde{t}^2+D}\exp\left(-\tilde{\chi}\right).
\ee
Properties of the solutions for the same four special cases discussed in the previous section are summarized in Table \ref{tab:disformal_beta_equal_zero} in Appendix \ref{appendix:table}.  These solutions are shown in the three right panels of Fig.~\ref{fig:four_background_solutions} for three different values of the only free parameter $D$.  The red curve in these panels corresponds to the solution of the complete equation (Eq.~\ref{disformal_no_beta}).

These results can be summarized as follows.  The evolution of the field at early times can be described analytically by linearizing the equation with respect to time and the field itself. Since the coefficient that appears in front of the time derivative in the damping term approaches zero at early times, it is possible to neglect it in this regime (note that this cannot be done in the quintessence model, for which this coefficient diverges at $\tilde{t}=0$).  The solution for the early universe is then a power law with a logarithmic slope equal to four, which is higher than in the quintessence case and thus, implies a slower evolution at early times.  Once this early stage is finalized, three different processes dictate the further evolution:
\begin{enumerate}
\item The two explicit functions of time that exist in the two terms in the right hand side of Eq.~\ref{disformal_no_beta} approach the quintessence value ($1/\tilde{t}$ and 1 respectively).
\item  The damping term increases with time.
\item  The term that forces the field to roll down the potential (second term on the right hand side of the equation) drops (i.e. the non-linear regime is reached).
\end{enumerate}
Three possible solutions exist depending on which of these processes is activated first, which in turn depends on the amplitude of $D$.  For small values of $D$ (red curve in the second panel from left to right of Fig.~\ref{fig:four_background_solutions}), the transition towards the quintessence limit occurs first (i.e. $T_b<T_a$ and $T_b<T_c$).  In this case the equation is transformed into the same equation that defines the quintessence model and thus the evolution continues following the solutions described in the previous section and in Table \ref{tab:quintessence} in Appendix \ref{appendix:table}.

For intermediate values of $D$ (red curve in the third panel from left to right of Fig.~\ref{fig:four_background_solutions}), the damping term dominates before the occurence of the transition to the quintessence regime.  In this case the acceleration of the field becomes negative and the field loses kinetic energy.  The solution becomes flat with a logarithmic slope equal to zero.

Finally, for very large values of $D$ (red curve in the fourth panel from left to right of Fig.~\ref{fig:four_background_solutions}), the evolution of the field stays in the disformal regime until the exponential function in the potential starts decaying.  The equation of motion becomes the equation of a free particle, whose solution has a logarithmic slope equal to one.  Once this happened, the field continues evolving unperturbed until the damping term dominates.  At this moment, the field decelerates and the slope becomes equal to zero.

\begin{figure}
  \includegraphics[width=0.48\textwidth]{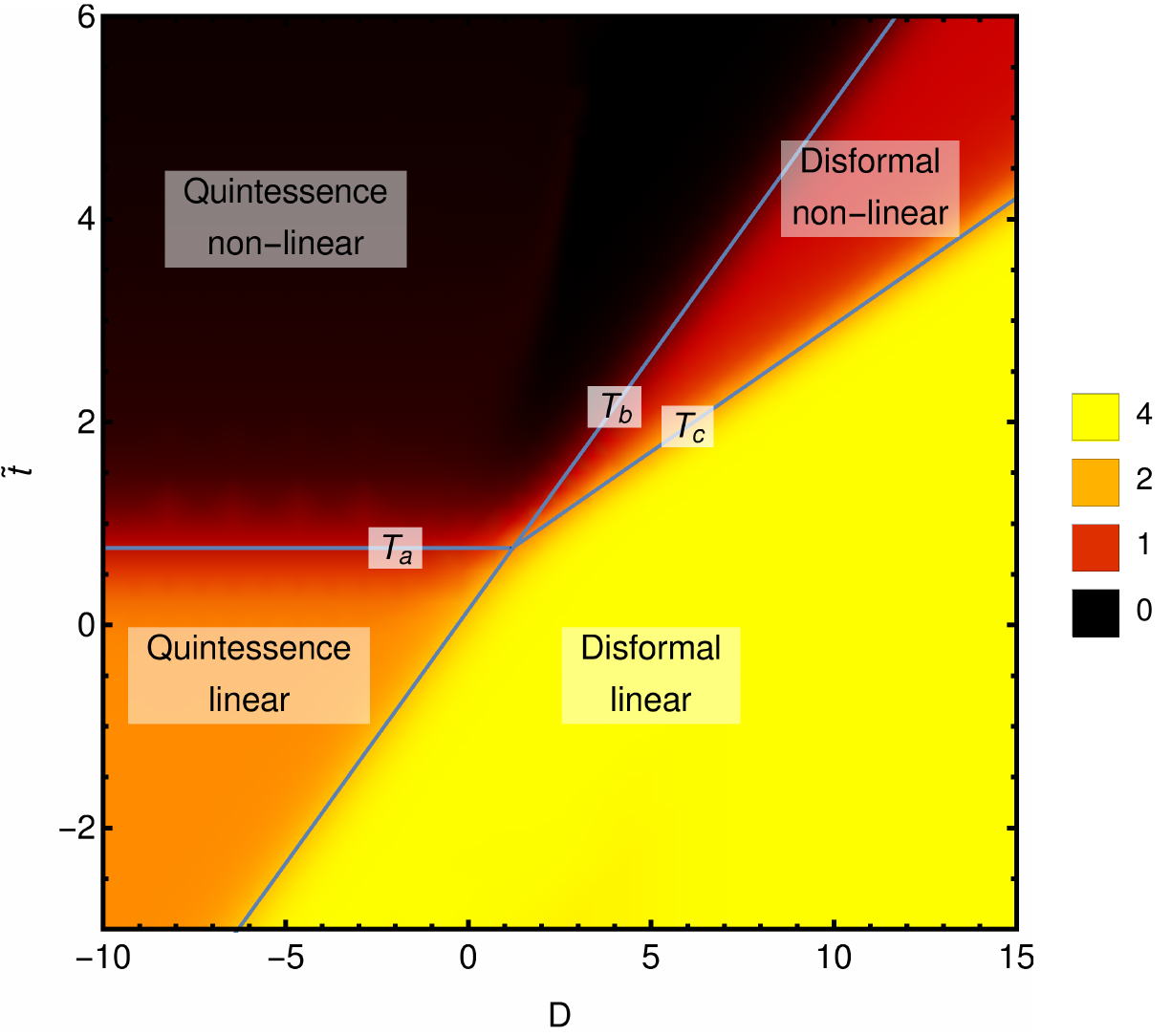}
  \caption{Colour coded is the logarithmic slope of the solution of the background Klein-Gordon equation for the disformal model with $\beta=0$ and an Einstein-de Sitter universe.  The three lines correspond to the three time scales defined in Table \ref{tab:definitions} (for the case that includes damping).}
  \label{fig:three_lines_background_solutions} 
\end{figure}

These apparently complex solutions can be summarized in a simple way by plotting their logarithmic slope as a function of $D$ and $\tilde{t}$, which we calculated numerically and show in Fig.~\ref{fig:three_lines_background_solutions}.  The transition between different regimes is given by the characteristic time scales defined in Table \ref{tab:definitions}.

\subsubsection{Disformal gravity with exponential disformal coupling}

Assuming $\beta$ (which is equivalent to $F$ in the parametrization studied here) is different than zero has two effects in the equation of motion (Eq.~\ref{eom_back}).  Firstly, it gives an additional dependence with the field $\tilde{\chi}$ to the coefficients that control the transition between the disformal and quintessence regimes. This new dependence converts the parameter $D$ into an effective parameter $\tilde{D}=D \exp(F\tilde{\chi})$.  Positive or negative values of $F$ will result in $\tilde{D}$ increasing or decreasing with time, which in turn, will delay or accelerate the transition to the quintessence regime (defined by $T_{\mathrm{b}}$).  Secondly, assuming $F\neq 0$ will add a new term to the equation of motion which will act as an additional damping term or external force on the field depending on the sign of $F$.

\begin{figure*}
  \includegraphics[width=\textwidth]{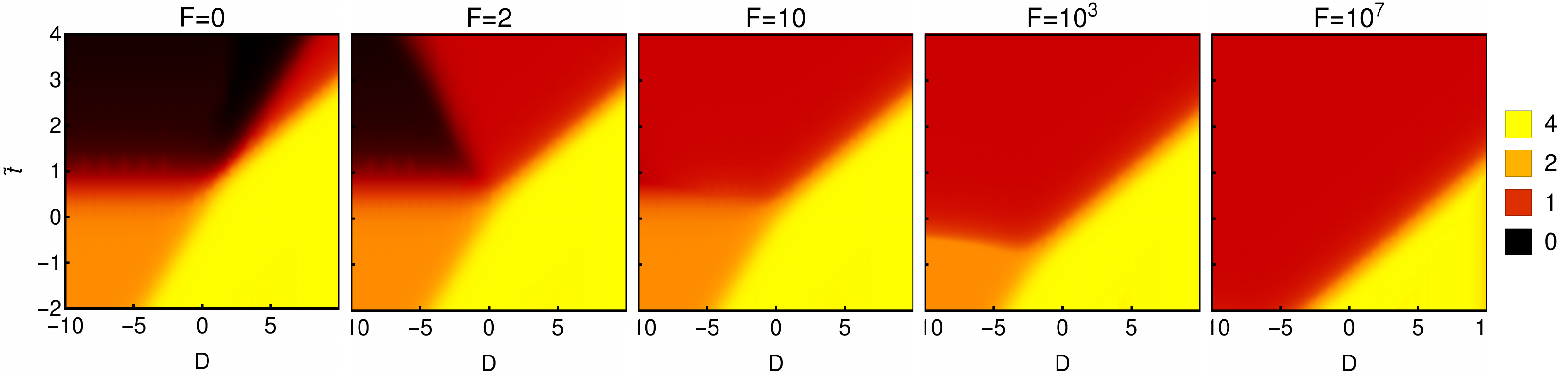}
  \caption{Color coded is the logarithmic slope of the solution of the background Klein-Gordon equation for the disformal model for different values of $F$ in an Einstein-de Sitter universe.}
  \label{fig:background_with_f}
\end{figure*}

In the case $F>0$, the parameter $\tilde{D}$ will grow exponentially with the value of the field.  This will decrease the time required to reach the transition to the non-linear regime as well as decrease the importance of the damping terms.  Since the new damping term decreases faster with time than the usual one, it will not have any impact on the overall shape of the solutions.  Figure \ref{fig:background_with_f} shows the logarithmic slope of the solution in the plane $($D$, \tilde{t})$ for different values of $F$.  For small values of $D$ and intermediate values of $F$ (second panel from the left), the damping term kicks in before the transition to the non-linear regime is reached and thus, the solution gets flat, with a logarithmic slope equal to zero.  However, the dependence of $\tilde{D}$ with $\tilde{\chi}$ will force the damping to decrease faster that the usual case, thus giving the chance to the potential term to resurge and change the logarithmic slope of the solution back to one.   Afterwards, all the terms on the right hand side will dissapear, which let the field evolve as a free particle, with a slope equal to one.  For large values of $F$, the transition to the non-linear disformal regime is faster than the transition to the linear quintessence regime, and so the slope of the solution has a direct transition from four to one.

To analyse the case $F<0$, it is convenient to re-write the equation of motion (Eq.~\ref{eom_back}) as follows:
\be
\partial_{\tilde{t}}^2\tilde{\chi} = 
-\frac{1}{4}\left[\frac{4\tilde{t} + FD \partial_{\tilde{t}}\tilde{\chi}}{\tilde{t}^2 + D\exp(F\tilde{\chi})}\right]
 \partial_{\tilde{t}}\tilde{\chi} + \frac{\tilde{t}^2}{\tilde{t}^2+D\exp(F\tilde{\chi})}\exp\left(-\tilde{\chi}\right). 
\ee
Here, it becomes evident that the condition for the first term of the right hand size to be negative (and thus, to act as a damping term instead of an external force) is
\be
4\tilde{t} + FD \partial_{\tilde{t}}\tilde{\chi} > 0.
\ee
By substituting the possible asymptotic limits discussed in Table \ref{tab:disformal_beta_equal_zero} (i.e. $\tilde{t}^4$, $\tilde{t}^2$, $\tilde{t}$ and $\log\left(\tilde{t}\right)$), we can see that the equation of motion will eventually become unstable for negative values of $F$.  In the particular regime in which $\tilde{\chi}\propto \tilde{t}^2$, the solution is unstable for all $\tilde{t}$.

\subsection{Disformal field dynamics:  perturbations}
\label{section:perturbations}

The model we are dealing with does not include conformal or explicit couplings.  However, the disformal coupling that we allowed for can generate perturbations in the field by itself.  The mechanism by which these perturbations originate is based on the fact that, thanks to the term $(1+\gamma^2\rho)^{-1}$ in the Klein-Gordon equation (Eq.~\ref{eq:EOM}), the field rolls down the potential $V$ at different rates in regions that have different densities (slower rate in higher density regions).  The shape of these perturbations can be summarized in a simple relation between the metric perturbation $\Psi$ and the scalar field $\phi$, which can be translated (after appropriate approximations) into a similar relation between the fifth force associated with the scalar field and gravity.

We can determine the exact form of field perturbations by re-writing the Klein-Gordon equation (Eq.~\ref{eq:EOM}) as follows:
\be
\nabla^2\phi =\frac{a^2}{c^2} \left[ \gamma^2\ddot{\phi} + \frac{1}{2}B_{,\phi}\dot{\phi}^2\right]  \delta\rho + \epsilon,
\label{jeans}
\ee
where we decomposed the density in a background value plus a (not necessarily small) perturbation
\be
\rho(\mathbf{x},t) = \rho_0(t) + \delta \rho(\mathbf{x},t)
\ee
and we defined
\be
\epsilon \equiv \left[ (1+\gamma^2\rho_0)\ddot{\phi} + 3H\dot{\phi} + \frac{1}{2}B_{,\phi}\rho_0\dot{\phi}^2 + V_{,\phi} \right]\frac{a^2}{c^2}, 
\ee
which is equal to zero in the background.  By substituting the factor $\delta\rho$ that appears in Eq.~\ref{jeans} with the corresponding value that can be obtained from the Poisson's equation for the metric perturbation
\be
\nabla^2\Psi = \frac{a^2}{2M_P^2} \delta\rho, 
\ee
we can relate the Laplacian of the scalar field and the metric perturbation:
\be
\label{laplacians}
\nabla^2\phi =  \xi \nabla^2\Psi + \epsilon, 
\ee
where we have defined
\be
\xi \equiv \frac{2M_P^2}{c^2}\left( B\ddot{\phi} + \frac{1}{2}B_{,\phi}\dot{\phi}^2 \right).
\ee
We now substitute all the dependences with the scalar field and its time derivatives with background values.  The term $\epsilon$ will disappear and we will end up with an equation that can be integrated and whose solution is
\be
\phi = \xi_0 \Psi + \phi_0 + \upsilon,  
\label{final_relation}
\ee
where $\upsilon$ is a solution of the Laplace equation, which, together with the background value of the field $\phi_0$, acts as an integration constant.  We will test this relation in the following sections by comparing results obtained from non-linear simulations which do not rely on approximations.  Since it will be used a posteriori, we show here the explicit form that the coefficient $\xi$ acquires when assuming an exponential potential as in Eqs.  \ref{eq:B} and \ref{defs_1}:
\be
\xi_0 = \frac{2 b_0}{H_0^2 c^2} \left(\ddot{\phi}_0 + \beta\frac{\dot{\phi}_0^2}{2 M_p}  \right) \exp(\beta \phi_0/M_p).
\ee
The sub-index zero in these expressions highlights the fact that these are background quantities.  The relation between the fields given by Eq.~\ref{final_relation} depends on the model parameter $b_0$ and $\beta$ and thus, breaks the degeneracy between parameters that enabled us to define $D$ and $F$ (Eqs.~\ref{defs_1} and \ref{defs_2}).  In the following section, we will study the consequences of breaking this degeneracy by simulating models that have the same background parameters $D$ and $F$, but different $b_0$.

An important characteristic of Eq.~\ref{final_relation} is that the sign of the coefficient $\xi_0$ depends in part on the sign of the second derivative of the field.  This means that at the moment in which the background field transitions towards the non-linear regime described in previous section (i.e.~the moment in which its second derivative becomes negative), the scalar field perturbations will be able to flip; their usual distribution will thus be inverted and local minima of the field will correspond to local minima in the density distribution.  The physical mechanism responsible for this phenomenon is related to the fact that the term $(1+\gamma^2 \rho)^{-1}$ in the Klein-Gordon equation (Eq.~\ref{eq:EOM}) increases the efficiency of the damping term in low density regions.  Thus, at the moment in which the damping term grows to the point in which it can affect the evolution of the field, the values in the halos overshoot that of the voids and inverts the sign of the perturbations.  Since the fifth force has an explicit dependence on the potential $V$ (and thus, on the scalar field itself), the flip in the perturbations will have a direct impact on the distribution of forces and the evolution of matter that is defined by it.

\subsection{The fifth force}  \label{sub:fifth_force}

The acceleration of a test particle in scalar-tensor theories can be found by studying the Jordan frame geodesics equations \citep{DBIgalileons, DisformalSymmetron}.  In the weak field limit of a general theory with a purely disformal coupling, the geodesic equation for a non-relativistic test particle is given by 
\begin{equation}
\ddot{\mathbf{x}} + 2H\dot{\mathbf{x}} + \left ( \boldsymbol{\zeta} \cdot \dot{\mathbf{x}} \right ) \nabla\phi  + \frac{\nabla \Psi}{a^2} + \frac{c^4}{2M_P^2}\frac{\xi}{g_{\phi}}\nabla\phi = 0, 
\label{eq:accel}
\end{equation}
where we have defined
\be
g_{\phi} \equiv 1-2BX
\ee
to simplify the notation and the 3-vector $\boldsymbol{\zeta}$ is a function of derivatives of $\phi$. Eq.~\ref{eq:accel} is equivalent to Newton's second law, where the acceleration of a body is proportional to the sum of forces acting on it.  The second term on the left hand side of the equation corresponds to a damping force induced by the expansion of the Universe; the third term is second order and the last two correspond to the Newtonian and fifth forces, which we define as:
\begin{align}
\label{def:force_1}
{\mathbf{F}}_\Psi & \equiv - \frac{\nabla\Psi}{a^2}, \\
\label{def:force_2}
{\mathbf{F}}_\phi & \equiv - \frac{c^4}{2M_P^2}\frac{\xi}{g_{\phi}} \nabla\phi.
\end{align}
These two force fields can be related to each other by taking into account the connection that exists between the scalar field and the metric perturbation discussed in the previous section.  As the version of that relation that does not rely on approximations applies to the Laplacian of the fields (Eq.~\ref{laplacians}), it is convenient to study the divergence of these force fields.  Thus, by taking into account definitions \ref{def:force_1} and \ref{def:force_2} and Eq.~\ref{laplacians}, we can write
\be
\nabla \cdot {\mathbf{F}}_{\phi} = \left( 1 - \delta d \right) \eta^2 \nabla\cdot{\mathbf{F}}_{\Psi},
\label{relation_divergences}
\ee
where the quantity
\be
\delta d \equiv \frac{g_{\phi}}{a^2 \xi^2\nabla\cdot{\mathbf{F}}_{\Psi} } \left[\nabla\left(\frac{\xi}{g_{\phi}}\right) \cdot\nabla\phi + \frac{\xi}{g_{\phi}} \epsilon\right]
\ee
is exactly zero in the background and we defined
\be
\eta^2 \equiv \frac{c^2}{2 M_P^2} \frac{a^2 \xi^2}{g_{\phi}}.
\label{def_eta}
\ee
The sign of $\eta^2$ depends exclusively on the sign of $g_{\phi}$, which must be positive for the theory to be stable. This is because the disformal transformation (Eq.~\ref{eq:coupling}) becomes singular when $g_{\phi} = 0$ (i.e~$\tilde{g}^\mu_\mu = 0$ and thus, the metric becomes not invertible), so this crossing must be avoided. Futhermore, \cite{DisformalScreening} showed that the evolution of the field will naturally avoid this singularity by progressively freezing the field before $2BX$ reaches unity. We can therefore assume $g_{\phi} > 0$, which results in $\eta^2$ being positive.

Eq.~\ref{relation_divergences} can be integrated after evaluating the coefficient $(1-d \delta)\eta^2$ in the background, which is the same as we did in the previous section when integrating Eq.~\ref{laplacians} to connect the field with the metric perturbations.  The end result is:
\be
\mathbf{F}_{\phi} =  \eta^2_0 \mathbf{F}_{\Psi} + \nabla\times \mathbf{k},
\label{relation_forces}
\ee
where the curl field $\mathbf{k}$ is an integration constant (in the sense that its divergence is zero) and we took into account that $\delta d$ is equal to zero in the background.  The properties of $\nabla\times\mathbf{k}$ are very well known in the context of Modified Newtonian Dynamics (MOND).  In particular, it has been shown that it is exactly zero for particular symmetries and that behaves at least as $r^{-3}$ for non-symmetric configurations \citep{1984ApJ...286....7B}.  Its effects in structure formation (in the context of MOND, which deals with a universe without dark matter) were studied in detail by \cite{2008MNRAS.391.1778L} (additional results associated with this paper can be found in \cite{llinares_thesis}).

The results described in this section and the previous one can be used to define two different simulation methods which will depend on two different approximations.  First, it is possible to neglect the effects induced by the term $\upsilon$ in Eq.~\ref{final_relation}.  Assuming also that $\xi_0$ is independent of the position (which we actually did to derive that equation), we can discretise the space and time derivatives that appear in the definition of the fifth force (Eq.~\ref{def:force_2}).  Thus, the fifth force can be calculated as a linear combination of values of $\phi$ in space and time.  We will discuss in the following section the impact that making these approximations has in the estimation of the scalar field by comparing with exact results obtained from cosmological simulations.  A companion paper will also contain a detailed estimation of how the error associated with these approximations translates into the predictions of observable quantities.

A different simulation approach that can be defined from these results consists in neglecting the curl term in Eq.~\ref{relation_forces} to find a relation between the Newtonian and fifth force fields.  This second case is equivalent to assume that all the gravitational effects can be condensed in an effective gravitational constant $G_{\mathrm{eff}} = G (1+\eta_0^2)$, which is assumed to be independent of the position.  Note that similar approximations were discussed in \cite{SaksteinD1} in the context of Solar System tests.

\begin{table*}[t!]
\centering
  \begin{tabular}{l | c c c c | c c | c | l }
  Model  & $v_0$ & $\nu$ & $b_0$ & $\beta$ & $D$ & $F$ & $v_0 \nu$ & Notes \strut \\
  \hline \strut
  GR  & --- & --- & --- & --- & ---& --- & --- & $\Lambda$CDM with Planck 2015 parameters \strut \\
  DDE & 3.055 & 0.4 & 1  & (-10,0,10) & 0.65 & $(-25,0,25)$ & 1.22 & Correct amount of Disformal Dark Energy \strut \\
  Fiducial & 1 & 1 & 1 & 0  & 1.3 & 0 & 1 & ---  \strut \\
  VF & 10 & 1 & 0.1 & 0  & 1.3 & 0 & 10 & Velocity flips in less than a Hubble time.  \strut \\
  Steep & $10^{-3}$ & $10^3$  & 0.01 &  0 & $1.3 \times 10^1$ & 0 & 1 & $V$ is steep $\Rightarrow$ fast transition to non-linear phase. \strut \\
  FF &  0.1 & 100  & 1 &  (-10,0,10) & $1.3\times 10^3$ & $(-0.1,0,0.1)$ & 10 & Field flips in less than a Hubble time. \strut
  \end{tabular}
\caption{\label{table:params} Model parameters used for the $N$-body runs. See Section \ref{section:nbody} for explanation.}
\end{table*}

\subsection{Achieving a repulsive fifth force}
\label{sub:analytic_repulsive}

The presence of non-canonical kinetic terms in the definition of a scalar tensor theory, can give rise to repulsive forces \citep{2004PhRvL..93r1102A}.  Since disformal gravity fulfils this condition when written in the Jordan frame, it may be worth investigating how repulsive forces can be achieved in this model.  The possibility of obtaining repulsive fifth forces is not only interesting from a theoretical perspective, but becomes relevant in the context of the discrepancy found by the Planck collaboration between the normalization of density perturbations $\sigma_8$ that can be inferred from the CMB and from lensing \citep{PlanckLensing}.  Although the measured discordance could be due to unknown biases or even statistical fluctuations \citep{PlanckDiscordance, 2017A&A...597A.126C, 2016MNRAS.459..971K}, these results could also be a signal of new physics.  A repulsive MG force will delay clustering with respect to GR and thus could help in reducing the tension.

The necessary condition for the fifth force to be opposite to gravity can be obtained from the relation between these two force fields provided by Eq.~\ref{relation_forces}.  However, since approximations where made when deriving this equation, a more appropriate starting point for this analysis is the divergence of this relation, for which we present an exact expression in Eq.~\ref{relation_divergences}.  This equation tells us that the only way in which the divergence of the two force fields can have different sign is by having 
\be
\delta d > 1.
\label{delta_large}
\ee
Since this quantity is equal to zero in the background, the disformal fifth force is parallel to gravity at order zero.  At first order in perturbations of the field, $\delta d$ takes the following form:
\be
\delta d \sim \frac{\left( 1 + \gamma^2\rho_0 \right)\ddot{\delta\phi} + 3H\dot{\delta\phi} + V_0\nu^2/M_P^2\delta\phi}{a^2 \xi^2 \nabla\cdot{\mathbf{F}}_{\Psi} }, 
\label{final_repulsive}
\ee
where we assumed $\beta=0$ for simplicity.  Since the sign of $\delta d$ depends on the sign of the perturbations, this quantity can certainly be positive.  It will also become larger than one at least in the specific redshifts in which $\xi$ changes sign and in regions where $\nabla\cdot{\mathbf{F}}_{\Psi}$ changes sign (i.e. in the transition between voids and over-dense regions).  So we can be certain that repulsive forces do exist in this model at least as a transient and in specific regions of space.  This result show that the approximations that we made to obtain Eq.~\ref{relation_forces} from Eq.~\ref{relation_divergences} may be important in specific situations and thus cannot be taken by granted in general.  This may be important in particular because voids and their outer limits, where $\nabla\cdot{\mathbf{F}}_{\Psi}$ approaches zero, were discussed in several opportunities as relevant probes of modified gravity \citep{llinares_thesis, 2015MNRAS.451.1036C, 2015JCAP...08..028B, 2017PhRvD..95b4018V, 2018MNRAS.475.3262F}.

An additional result that comes from Eq.~\ref{final_repulsive} is that the fifth force has an explicit dependence with the potential $V$, and thus with the absolute value and sign of the perturbations.  This may be important since we showed in Section \ref{section:perturbations} that the scalar field perturbations are proportional to the metric perturbations (Eq.~\ref{final_relation}) and that they can flip following the sign of the coefficient $\xi_0$.  So the moment in which the field perturbations flip, will be associated with a change in the amplitude of the fifth force.

\section{\texorpdfstring{$N$}{section:nbody}-body Simulations}
\label{section:nbody}

We summarize in this section technical aspects of the 3D cosmological simulations that we run to both confirm the results presented above in a realistic set up and quantify the impact that the fifth force has in the matter distribution in the non-linear regime.  This section also describes in detail our motivation for choosing the particular set of model parameters that we simulated.

\subsection{Set up of the simulations}

The simulations were run with the modified gravity N-body cosmological code \texttt{Isis} \citep{Isis}, which is based of the particle mesh code \texttt{Ramses} by \cite{Ramses}.  The code includes a solver for the non-linear MG elliptic equations that can be obtained after assuming the quasi-static approximation.  However, since in disformal gravity the time derivatives play a central role in both the Klein-Gordon and geodesics equations (Eqs.~\ref{eq:EOM} and \ref{eq:accel}), the validity of this approximation is not guaranteed.  Thus, we made use of the non-static solver of \texttt{Isis} \citep{Nonstatic,DisformalSymmetron}, which relies on a non-linear hyperbolic solver that can take into consideration time derivatives of the background and perturbed fields.

To be consistent with the simulation code \texttt{Isis/Ramses}, we will use the supercomoving time $\tau$, which relates to cosmic time $t$ through $d\tau = dt/a^2$ \citep{supercomoving}.  We will denote derivatives with respect to this new time with a prime.  We will also work with the following normalized scalar field
\begin{equation}
\chi \equiv \frac{\phi}{M_\mathrm{pl}}.
\end{equation}
The code variables associated to the time derivatives $\dot \phi$ and $\ddot \phi$ are 
\begin{align}
q &\equiv a \chi ', \\
q' &= a' \chi ' + a \chi ''.
\end{align}

We performed $N$-body simulations with $256^3$ particles in a cubic box with a comoving side length of 256 Mpc$/h$.  The background cosmology is assumed to be the same for all the runs and given by Planck 2015 best fit $\Lambda$CDM parameters \citep{Planck2015}: $(H_0, \Omega_{\Lambda}, \Omega_m, \sigma_8) = (67.74$ km/s/Mpc, 0.6911$, 0.3089, 0.8159$).  The assumption behind this selection for the background expansion is that the energy of the scalar field is compatible with a cosmological constant with the value required by observations.  Thus, from the numerical point of view, taking into account the energy of the scalar field is equivalent to adding a cosmological constant.  We note that not all the models that we simulated possess this property.  However, in this paper, rather that finding the best fit disformal parameters, we are interested in looking for new phenomenology which may provide us with novel ways of looking at the data.  This may potentially lead to a detection of deviations from GR in the data, which does not necessarily correspond exactly to the model studied here, but that share observable signatures with it.  Taking into account the energy of the scalar field will complicate the analysis, but will not provide additional information on the disformal effects associated with the perturbations.

The initial conditions for the N-body particles were generated assuming that the impact of the scalar field at high redshift is negligible.  Thus, all the simulations use the same initial conditions, which were generated with the Zeldovich approximation with the code \texttt{Grafic2} \citep{Grafic2}.  These sets of initial conditions share not only the power spectrum, but also the phases.  By doing this, we ensure that differences found between the various simulated models are induced by the presence of a fifth force and not the initial particle distribution.

\begin{figure*}
  \begin{minipage}[t]{0.48\textwidth}
    \includegraphics[width=\columnwidth]{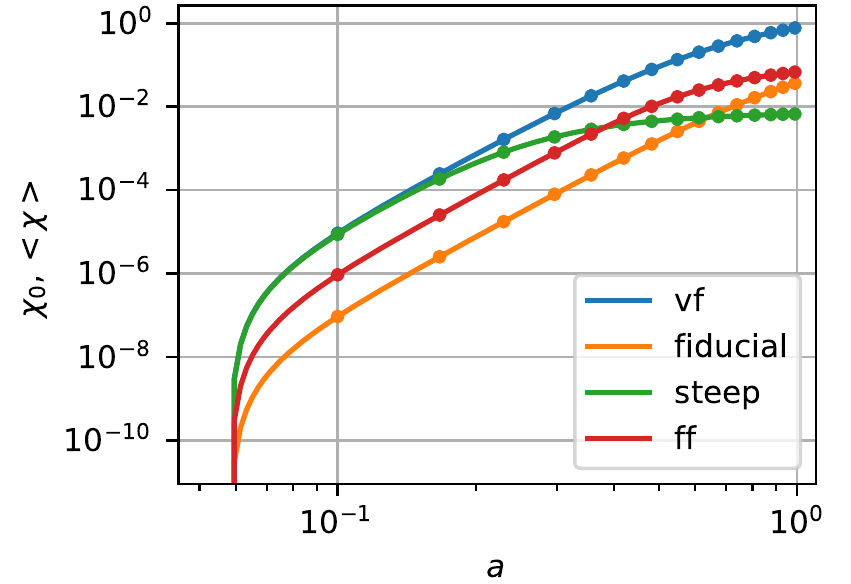}
  \end{minipage}
\hfill
  \begin{minipage}[t]{0.48\textwidth}
    \includegraphics[width=\columnwidth]{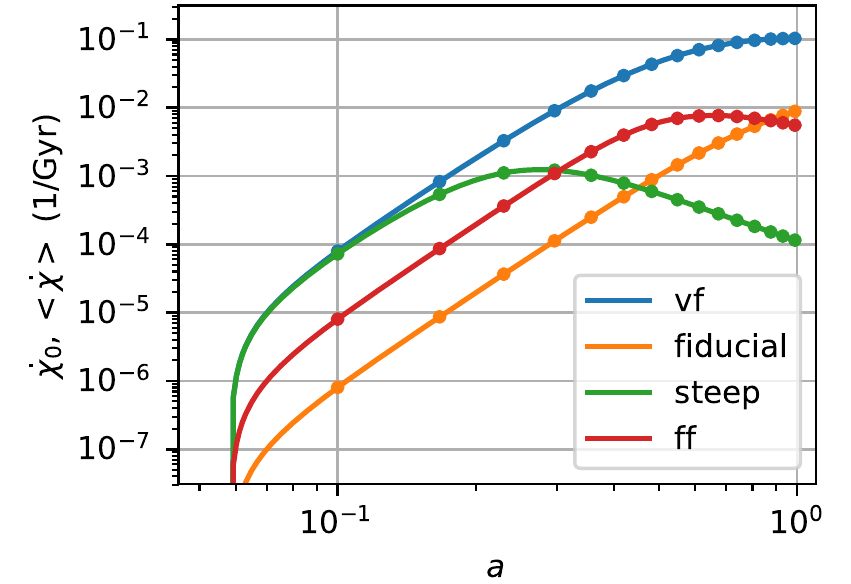}
  \end{minipage}
  \caption{\label{fig:comparison_background} Background evolution of the scalar field (left) and its time derivative (right) for the simulated models as a function of the expansion factor $a$.  The continuous curves are Runge-Kutta solutions of the order zero Klein-Gordon equation and the points are the mean values obtained from the non-linear simulations in slices that pass through the centre of the box.  Results of the DDE run are very similar to those of the Fiducial run, so we exclude them to avoid overcrowding the plots.}
\end{figure*}

The codes \texttt{Ramses} and quasi-static \texttt{Isis} include adaptive mesh refinements (AMR), which means that they can increase the resolution locally as required by the complexity of the solutions.  However, the non-static version of \texttt{Isis} does not include this technique.  Thus, our analysis will only be valid up to the Nyquist frequency of the domain grid which covers the whole simulation box and has 256 nodes per dimension.  The three-dimensional particle data was output at nine different snapshots, at $z=2.33$, 1.00, 0.43, 0.25, 0.11, 0.081, 0.053, 0.026, and 0. In addition, the code outputs all the available fields (density, metric perturbation, scalar field and its derivatives) in a two-dimensional slice that crosses the center of the box at 200 different points in time, ranging from $z=16$ to $z=0$.

\subsection{Simulated models}

Table \ref{table:params} lists the model parameters that we chose for the simulations.  We also included the derived background parameters $D$ and $F$ defined in Section \ref{background_evolution} and the product $v_0\nu$ which will be useful to interpret results in the following sections.  Note that the background parameters $D$ and $F$ were original defined for an Einstein-de Sitter universe, so they must be taken only as indicative.  The slope of the disformal coupling $\beta$ is set to zero for all the runs except the models DDE and FF, for which we made two additional runs with $\beta=\pm 10$.  A brief explanation of the motivation for each set of parameters follows.

In the Fiducial run, all three free parameters of the model are set to one.  These values give a background evolution that is very close to that of the $\Lambda$CDM model.  To confirm this, we also run the Disformal Dark Energy (DDE) simulation with parameters that were specifically tuned to recover the $\Lambda$CDM expansion rate.  The parameters of this simulation are such that the potential $V(\phi)$ gives the observed amount of Dark Energy at redshift zero and at the same time stays within constraints obtained in the linear regime by \cite{MifsudPlanck}. We find that their upper limit for the coupling, $B_0 = \left( 0.2\, \mathrm{meV} \right)^4$, is equivalent to $B_0 \approx 1/ \left( H_0^2 M_\mathrm{pl}^2 \right)$ in the units used in this work. Consequently, using a dimensionless coupling $b_0 = \mathcal{O}\left(1\right)$ will give a model that is within the constraints given in that study. The initial potential, $v_0 = 3.055$, was found with a shooting algorithm, following \cite{MifsudPlanck}. Combining this with a small value for $\nu$ (i.e. a flat potential) results in $V \sim 3 H_0^2 M_\mathrm{pl}^2$ today, and hence gives rise to a dark energy component similar to a cosmological constant.  Most of the results that come out of these two simulations (Fiducial and DDE) are similar. Thus, we will show only results of the Fiducial run and point out differences when needed.

The parameters of the Field Flipping (FF) and Velocity Flipping (VF) models were chosen to present a broad spectrum of phenomenology. The FF simulation enters the non-linear background phase discussed in Section \ref{disformal_background_constant_coupling} before redshift zero.  The analysis presented in Section \ref{section:perturbations} shows that this will induce a flip in the scalar field perturbations.  Once this happens, the distribution of the field perturbations will contradict the usual profile for a coupled scalar field and will associate high density regions with local maxima in the field.  In the VF case, a Hubble time is enough to get the flip in the time derivative of the field, but not in the field itself.  The parameters of this model were chosen to study how the degeneracy that exists between background parameters $D$ and $F$ can be broken by looking at the perturbations in the fields.  Thus, the simulation shares the values of $D$ and $F$ with the Fiducial simulation, however, it has a different value of $b_0$.

The Steep model was run using parameters that correspond to a steep potential $V$, and thus undergoes an early transition towards the non-linear regime (i.e. corresponds to a small value $T_c$).  The aim of this simulation is to analyse the consequences of a field that is active only at high redshift and is damped afterwards.

\begin{figure*}
\begin{center}
    \includegraphics[width=0.97\textwidth]{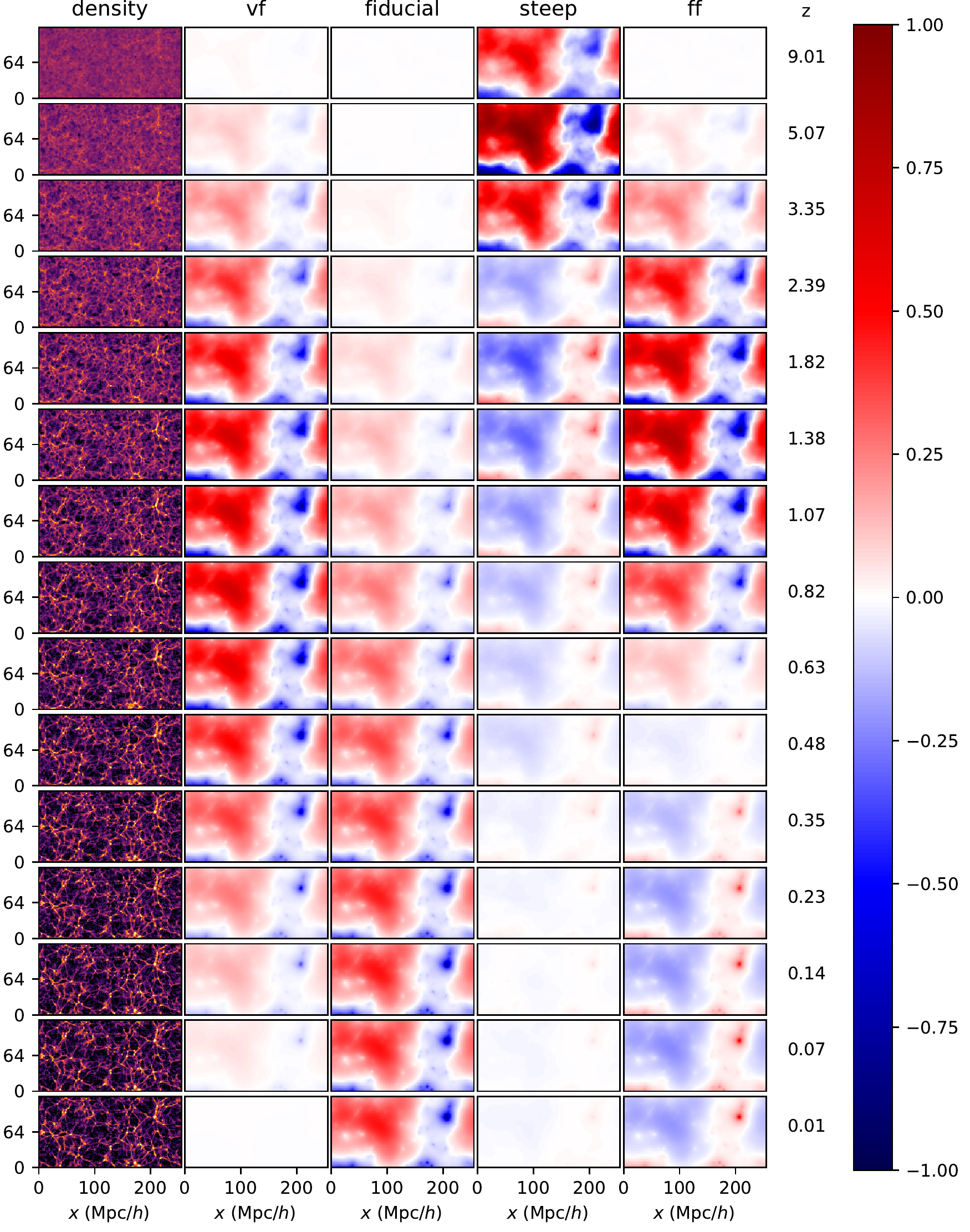}
\end{center}
\caption{\label{fig:time_evolution_delta_chi} The first column shows the evolution of the density distribution.  The rest of the columns show the time evolution of scalar field perturbations for different models. The colours correspond to the scalar field perturbations normalized to the maximum reached in each model, which means that the values go from -1 (dark blue) to +1 (dark red).  The scaling of the colours is symmetric-log.  The numbers next to the color bar correspond to the redshift of each row.  For reasons of space, we show only the bottom half of each slide.}
\end{figure*}

\section{Simulation results: properties of the fields}
\label{sec:simulation_results_field}

We discuss in this section several aspects of the distribution of the simulated fields as well as a comparison with the analytic estimations presented in Section \ref{section:analytical}.

\subsection{Background evolution of scalar field}
\label{background_field}

Given that the simulations track the scalar field $\chi$ rather than its perturbations $\delta\chi$ on a background $\chi_0$, it is worth asking if the mean value of the simulated scalar field agrees with the background value that can be calculated, for instance, with a Runge-Kutta solver.  We show such a comparison in Figure \ref{fig:comparison_background}.  The lines in the left and right panels are Runge-Kutta solutions for the background scalar field and its time derivative respectively.  The points are the mean values obtained from the N-body simulations in slices that pass through the centre of the box.  The abrupt decline towards zero at high redshift is related to the fact that the initial conditions are not given at $a=0$, but at the starting redshift of the simulations, when we assumed that the field is equal to zero.  Both solutions agree very well, which is a confirmation of a reliability of the code.

\begin{figure}
  \includegraphics[width=\columnwidth]{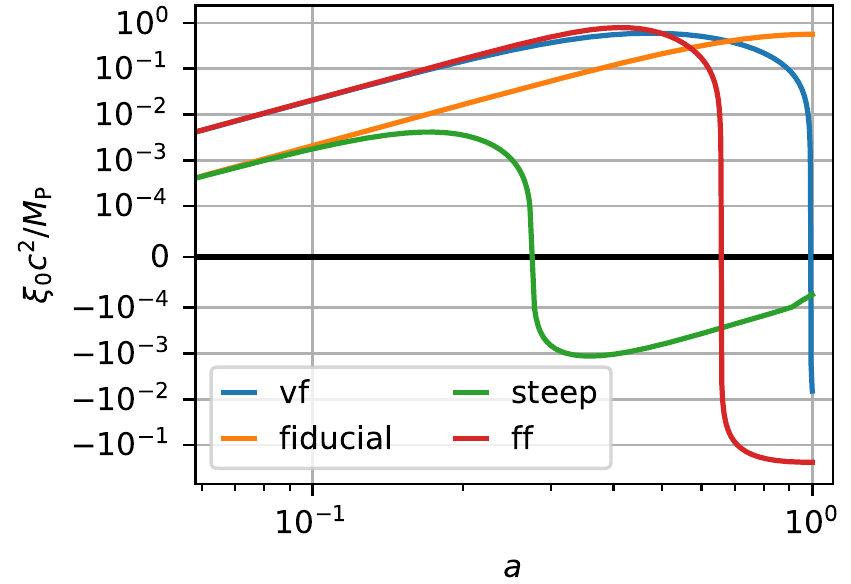}
  \caption{\label{fig:coefficient} Time evolution of the coefficient $\xi_0$ that relates metric perturbations $\Psi$ with the scalar field $\chi$ (Eq.~\ref{final_relation}) for the simulated models.  The DDE model is similar to the Fiducial one, and thus not shown to avoid overcrowding the plot.  The sharp transitions between positive and negative values are produced by the symlog scaling that we used for the vertical axis.  See section \ref{sec:fieldresults} for explanation.}
\end{figure}

\subsection{Qualitative behaviour of scalar field perturbations}\label{sec:fieldresults}

Figure \ref{fig:time_evolution_delta_chi} shows the time evolution of scalar field perturbations as found in 2D slices that pass through the center of the 3D box.  Different rows correspond to different redshifts (shown to the left of the color bar).  The first column shows the density distribution, while the other four correspond to the perturbation in the scalar field.  In this section we are only interested in a qualitative description of the effects associated to the disformal coupling, thus we normalized the perturbations with the maximum value reached in each model separatelly.  When using this normalization, the perturbations lie always between minus one and one.  A quantitative description will be presented in following sections.  Since the Fiducial and DDE models have almost identical evolution, we show only one of these two models.

The panels show that scalar field perturbations can do more that simply grow with time as happens in other scalar tensor theories.  The only model that has a monotonic growth of perturbations is the Fiducial one.  In the VF model, the perturbations grow until redshift of about $z=0.82$ and then wash out until being completely absent at redshift $z=0$.  In the Steep model, the perturbations grow very rapidly at early times, reaching their maximum at $z \sim 5$.  Then the field flips, developing maxima in the position of the halos and minima in the voids.  After the flip occurs, the perturbations washed out as happened in the VF model.  Finally, the FF model has a behaviour which is similar to the Steep model:  a maximum of the perturbations at $z\sim 1.38$, followed by a flip at $z\sim 0.48$.  In this particular case, the perturbations continue being almost constant until $z=0$.

The solutions shown in these panels seem very complex, but can be easily explained with the relation between the scalar field $\phi$ and the gravitational potential $\Psi$ that we discussed in Section \ref{section:perturbations} (Eq.~\ref{final_relation}).  Assuming that the integration constant $\upsilon$ in that equation is zero, we find that both fields are related through the parameter $\xi_0$, whose time dependence is shown in Figure \ref{fig:coefficient} for the same models presented in Figure \ref{fig:time_evolution_delta_chi}.  Two pairs of models have identical behaviour at high redshift.  This has the consequence that the early evolution of $\xi_0$\footnote{We derived this special case by taking into account the early redshift solution (Eq.\ref{eq:early_disf}) together with the definitions provided in Section \ref{background_evolution} and the definition of $\xi_0$.}:
\begin{multline}
\xi_0(t\ll 1) = \\
\frac{3 M_p H_0^2}{c^2}t^2\left[\frac{v_0\nu}{4} + \frac{M_PH_0^4}{85} \frac{v_0^2\nu^2\beta}{b_0}t^4 \right] \exp(\beta \phi_0/M_p)
\end{multline}
 depends on the product $v_0\nu$, which is the same for the two pairs of models (Fiducial, Steep) and (VF, FF) (see specific values in Table \ref{table:params}).  This degeneracy is broken at later times, and thus models that agreed at high redshift depart from each other later on.  The reason for this degeneracy to be absent in different columns of Figure \ref{fig:time_evolution_delta_chi} is that the normalization chosen for these panels is different for different models.  Finally, figure \ref{fig:coefficient} also shows that $\xi_0$ becomes larger for smaller values of $b_0$.  This may seem to contradict the fact that the solutions provided in Section \ref{section:analytical}, which depend only on $D$, are independent of $b_0$.  However, one must take into account that there is an additional factor $b_0$ in the definition of $\tilde{t}$ used in that section.

Comparison of Figures \ref{fig:time_evolution_delta_chi} and \ref{fig:coefficient} unveils the close relation that exists between the evolution of the perturbations and $\xi_0$.  The maxima in the perturbations agree with the maxima in $\xi_0$.  Also the moment in which the flips occur in the field agree with the change in the sign of $\xi_0$.

The fact that the scalar field perturbations shown in Figure \ref{fig:time_evolution_delta_chi} go back and forth in the FF simulation may open the question of the possibility of obtaining sustained oscillations in the field perturbations.  However, these changes in the perturbations are associated with transitions between different regimes in the background evolution of the field, which are condensed in the evolution of $\xi_0$.  The analysis presented in Section \ref{background_evolution} shows that after the second oscillation occurred, the perturbations cannot do more than to approach zero.

\begin{figure}
  \includegraphics[width=\columnwidth]{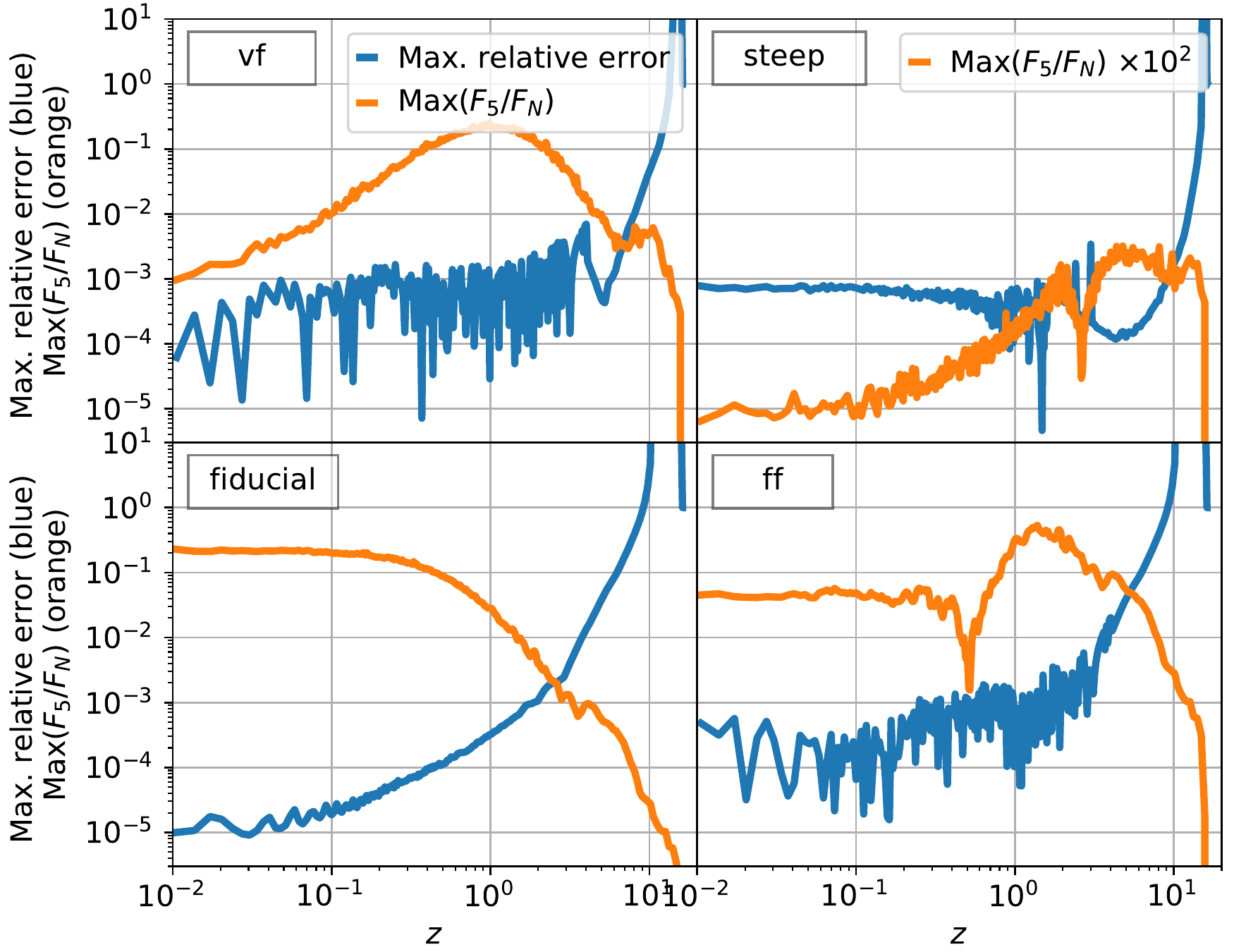}
  \caption{\label{fig:phi_psi}  The blue curves correspond to the maximum relative difference between the scalar field values obtained from the N-body simulations and the ones we obtained from the gravitational potential by assuming $\upsilon$ is zero in Eq.~\ref{final_relation}.  The orange curves are the maximum ratio between the fifth force ${\mathbf{F}}_\Psi$ and gravity ${\mathbf{F}}_\phi$.  We show results from 2D slices that pass through the center of the 3D box.  See Section \ref{section:testing_phi_psi} for explanation.}
\end{figure}

\begin{figure*}
\begin{center}
    \includegraphics[width=0.97\textwidth]{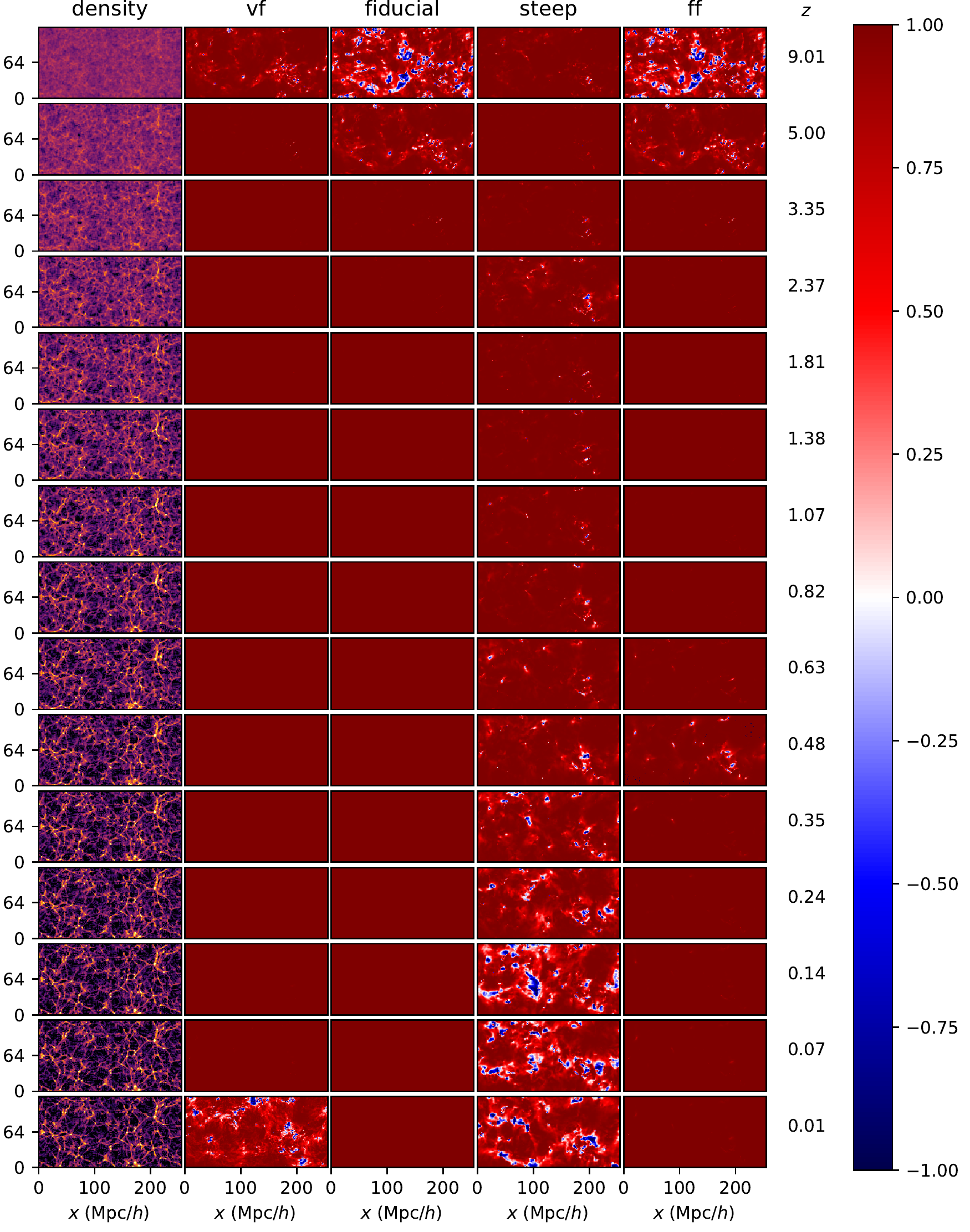}
\end{center}
\caption{\label{fig:time_evolution_angle_between_forces} Cosine of the angle between Newtonian and fifth force as a function of time and simulated model.  Dark red and blue correspond to force fields that are parallel or antiparallel respectively.  The numbers to the left of the colour bar correspond to the redshift of each row.}
\end{figure*}

\begin{figure}
\begin{center}
    \includegraphics[width=0.49\textwidth]{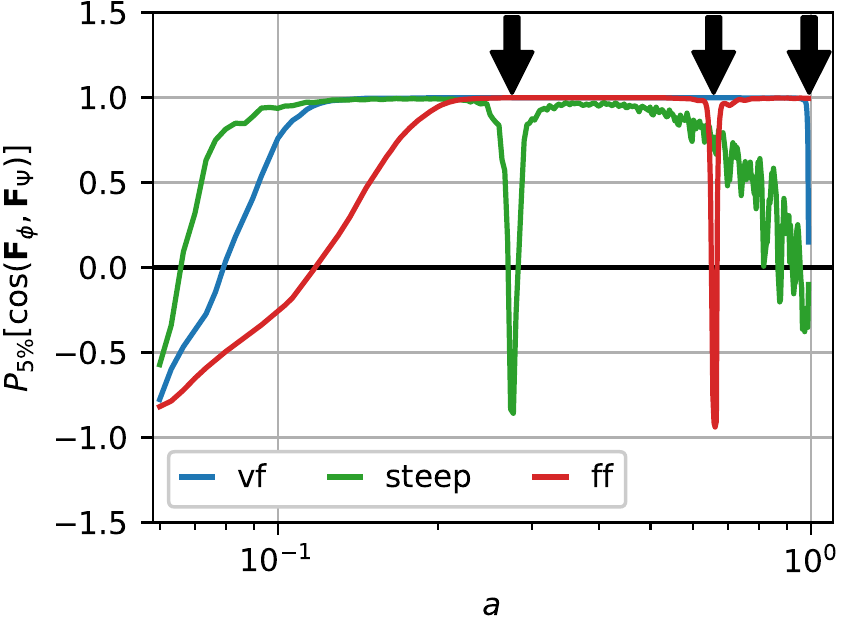}
    \caption{\label{fig:percentile_FdF} Percentile five of the distribution of the cosine of the angle between the Newtonian and fifth force as found in the 2D slides of our 3D simulations.  We show only the models for which we expect repulsive forces associated to the zeros of $\xi_0$:  VF, FF and Steep.  The vertical arrows show the zeros of $\xi_0$ for these models.}
\end{center}
\end{figure}

\subsection{Testing analytic predictions for field perturbations}
\label{section:testing_phi_psi}

A more quantitative comparison between the the scalar field that we extracted from the self-consistent N-body simulations and the predictions that we can obtain by assuming that $\upsilon=0$ in Eq.~\ref{final_relation} is shown in Figure \ref{fig:phi_psi}.  The blue curves are the maximum relative difference between these two fields as found in 2D slices that pass through the centre of the box and as a function of redshift.  Different panels correspond to different simulations.  The plots show that there is a very good agreement between the exact field provided by the simulations and our prediction: differences are below 0.1\% at all times after $z=1$.  The larger differences that occur at high redshift are related to the fact that we did not take into account perturbations of the scalar field in the initial conditions for the simulations (instead, we approximated them by zero).  Thus, there is a transient in which the field evolves from zero to a field that has a power spectrum compatible with Eq.~\ref{final_relation}.

These high redshift differences can be reduced by choosing more accurate initial conditions (for instance, generated as a realization of a linear power spectrum).  However, given that they occur at a moment in the history of the Universe when the fifth force is negligible, is it likely that they will not affect the matter distribution.  We confirmed this by plotting the maximum ratio between the fifth force ${\mathbf{F}}_\phi$ and gravity ${\mathbf{F}}_\Psi$ on the 2D slices as a function of time (orange curves in Figure \ref{fig:phi_psi}).  As expected, the fifth force is sub-dominant at high redshift.  By the moment in which the fifth force goes above 1\% of the Newtonian force, the error in our prediction of the scalar field is already below 0.1\% and continues decreasing from there.

The good agreement that we found between exact and approximate solutions shows that it does make sense to run disformal gravity cosmological simulations by using Eq.~\ref{final_relation} instead of a hyperbolic solver (as described at the end of Section \ref{section:perturbations}). This will largely reduce the overhead associated with the modified gravity solver and will make disformal simulations competitive in terms of speed.  The impact of this approximation in the predicted power spectrum of density perturbations will be presented in a companion paper.

\begin{figure*}
  \includegraphics[width=\textwidth]{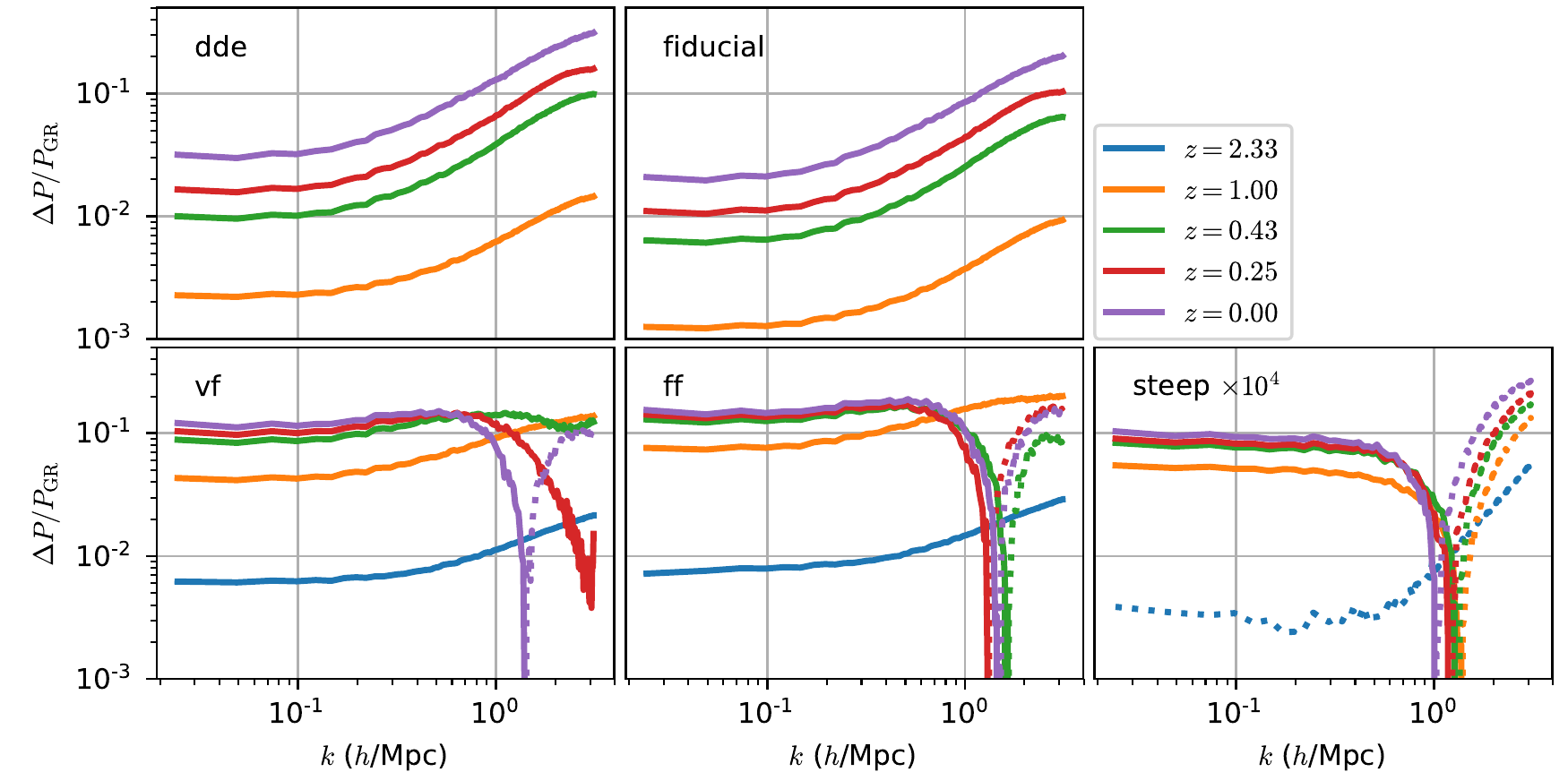}
  \caption{\label{fig:redshift_pk}  Relative difference between MG and GR dark matter power spectrum from all the simulations.  The values obtained from the Steep simulation are much smaller than the scale used for the other models, and thus we multiplied them by $10^4$ for presentation purposes only.}
\end{figure*}

\subsection{Direction of the fifth force}

In section \ref{sub:analytic_repulsive}, we discussed a sufficient condition for the presence of repulsive fifth forces (i.e. opposite to gravity), which is condensed in Eqs.~\ref{delta_large} and \ref{final_repulsive}.  Given that our simulations track the evolution of the scalar field without making any approximations or simplifications in the equations, it is possible to use the simulated force field to confirm if repulsive forces do exist in a realistic set up.  Figure \ref{fig:time_evolution_angle_between_forces} shows the cosine of the angle between the Newtonian and fifth forces in the same slices and for the same redshifts and simulations we presented in Figure \ref{fig:time_evolution_delta_chi}.  Dark red and blue correspond to parallel and antiparallel forces and the numbers to the left of the colour bar make reference to the redshift of each slice.

The panels show that all the simulations produce repulsive forces at high redshift, however these occur during the warm up phase of the simulation in which the scalar field transitions from the initial condition ($\phi = 0$) to a distribution that is consistent with the density field (see discussion in the previous section).  Moving forward in time, we see that repulsive forces appear only at specific redshifts.  These redshift values are consistent with the discussion presented in Section \ref{sub:analytic_repulsive}, where we found that repulsive forces will occur in the zeros of the function $\xi_0$.  This is confirmed in Figure \ref{fig:percentile_FdF}, where we show the fifth percentile of the distribution of the angle between the forces as a function of scale factor for the 2D slices.  The vertical arrows are the zeros of $\xi_0$, which are consistent with the moments in which the forces become antiparallel.

\section{Simulation results: Impact of the fifth force on the matter distribution} \label{sec:matterresults}

We describe now the effects that the presence of the fifth force has on the distribution of matter.  We do this by studying differences between the power spectrum of dark matter density perturbations of the GR and MG simulations.  We do firstly the analysis for the models that have $\beta=0$ and discuss a posteriori the differences induced by adding a dependence of the coupling with the scalar field through $\beta$.

\subsection{Power spectrum of models with $\beta=0$}

Figure \ref{fig:redshift_pk} shows the time evolution of the relative difference between the MG and GR power spectra for the models that have a constant disformal coupling (i.e. with $\beta=0$).  Different panels correspond to the different models that we simulated and different lines within each panel to different redshifts.  In the Steep run, the evolution is very close to GR, and thus we multiplied these curves by $10^4$.  As the curves have positive and negative values, we show their absolute values and highlight the negative parts with dotted lines.

The models can be divided in two different categories: those in which the fifth force produces a monotonic increase of power with respect to GR (models DDE and Fiducial) and those that have a monotonic increase until some specific redshift and then continue increasing at large scales, but have a slower evolution than GR at small scales (models FF, VF and Steep).

\begin{figure}
  \includegraphics[width=\columnwidth]{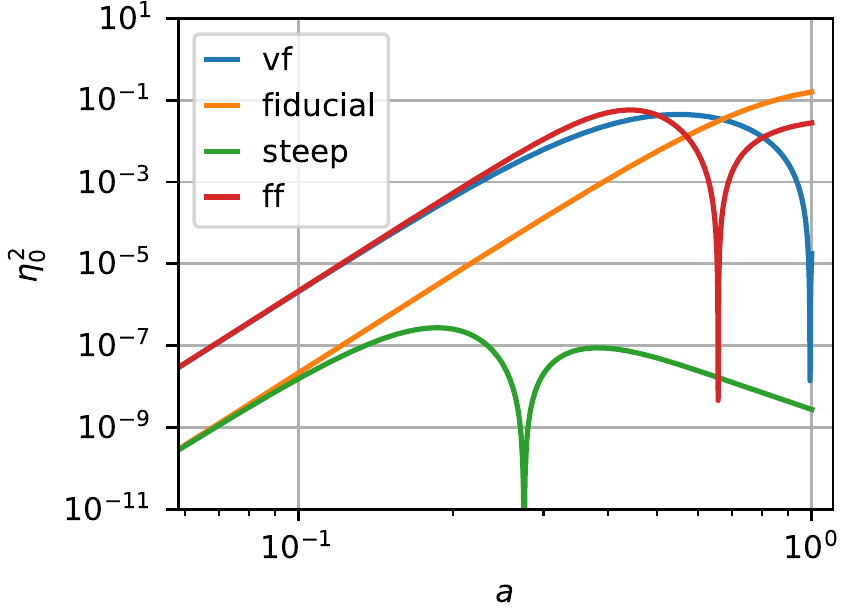}
  \caption{\label{fig:coefficient_eta} Time evolution of the dimensionless coefficient $\eta^2$ that relates Newtonian and fifth force fields when calculated with background quantities for the simulated models.    The DDE model is similar to the Fiducial one and thus, is not shown to avoid overcrowding the plot.  The discontinuities in the derivatives are not real, but produced by the log scaling.}
\end{figure}

Some insight on the phenomenology associated to these curves can be gained by comparing the amplitude of the fifth force with the gravitational force.  By assuming that $\delta d$ is second order in Eq.~\ref{relation_divergences}, we can obtain Eq.~\ref{relation_forces}.  If we further assume $\nabla\times\mathbf{k} \ll 1$, we find that the amplitude of the fifth force for a particular set of parameters depends exclusively on the background quantity $\eta_0^2$, which we show in Figure \ref{fig:coefficient_eta} for all the simulated models as a function of the expansion factor $a$.  Since the values of the DDE run are very close to those of the Fiducial run, we excluded them from this plot.  The discontinuities in the derivatives of these curves are not real, but related to the log scaling of the vertical axis.  However, these discontinuities do have a physical meaning, since they occur at moments when the parameter $\xi_0$ changes sign (see Figure \ref{fig:coefficient}).

From these curves, we can see that the pairs of models $M_1$ = (Fiducial, Steep) and $M_2$ = (VF, FF) have identical forces at high redshift and so it is expected that their early time evolution will be identical.  The fact that $\eta_0^2$ of the models $M_2$ is more than two orders of magnitude larger than that of the modes $M_1$ is responsible for the large differences found in Figure \ref{fig:redshift_pk}:  models $M_2$ have a much faster evolution, reaching differences with respect to GR of more than 10\% at large scales.  Note that the model FF has a smaller coupling constant than the Fiducial model.  Naively, one will expect that a smaller coupling constant will be associated to a slower evolution, however, the particulars of the definition of $\eta_0^2$ (in particular, the functions $\xi$ and $g_{\phi}$) give stronger forces in the FF case.

Comparison between the Steep and Fiducial values of $\eta_0^2$ show that even if both models share the same force fields at high redshift, they depart from each other later on.  In the Steep simulation, the forces stay always close to zero and thus differences in the power spectrum with respect to GR are minimal.

Figure \ref{fig:coefficient_eta} can also help us understand why the power spectrum of the MG simulations goes below the GR values at low redshifts.  In particular, it is possible to see that the moment in which the $\Delta P/P_\mathrm{GR}$ starts moving back to zero corresponds to the moment in which $\eta_0^2$ starts decreasing and approaches zero.  Also, this change in the behaviour of the power spectrum occurs at small scales, where the structure of the halos dominates the signal.  These facts are consistent with a decrease of power in MG simulations induced by an expansion of the halos when the fifth force disappears and the kinetic energy of the dark matter particles dominates.

Finally, we would like to point out that late evolution of the models DDE and Fiducial is almost scale independent.  After an initial shape is given to the difference between MG and GR, that shape is almost conserved later on.  Departure from scale invariance evolution of these curves occurs close to the Nyquist frequency of the simulations, and thus higher resolution simulations should be run in order to confirm this result.

Note that our analysis based on $\eta_0^2$ does not take into account the back-reaction that the presence of the fifth force has on the metric perturbations (through changes in the density distribution).  However, this effect was taken into account in the simulations, which are fully self-consistent.

\subsection{Response of the power spectrum to changes in $\beta$ \label{sec:powerspec_p}}

\begin{figure*}
  \centering
  \includegraphics[width=\textwidth]{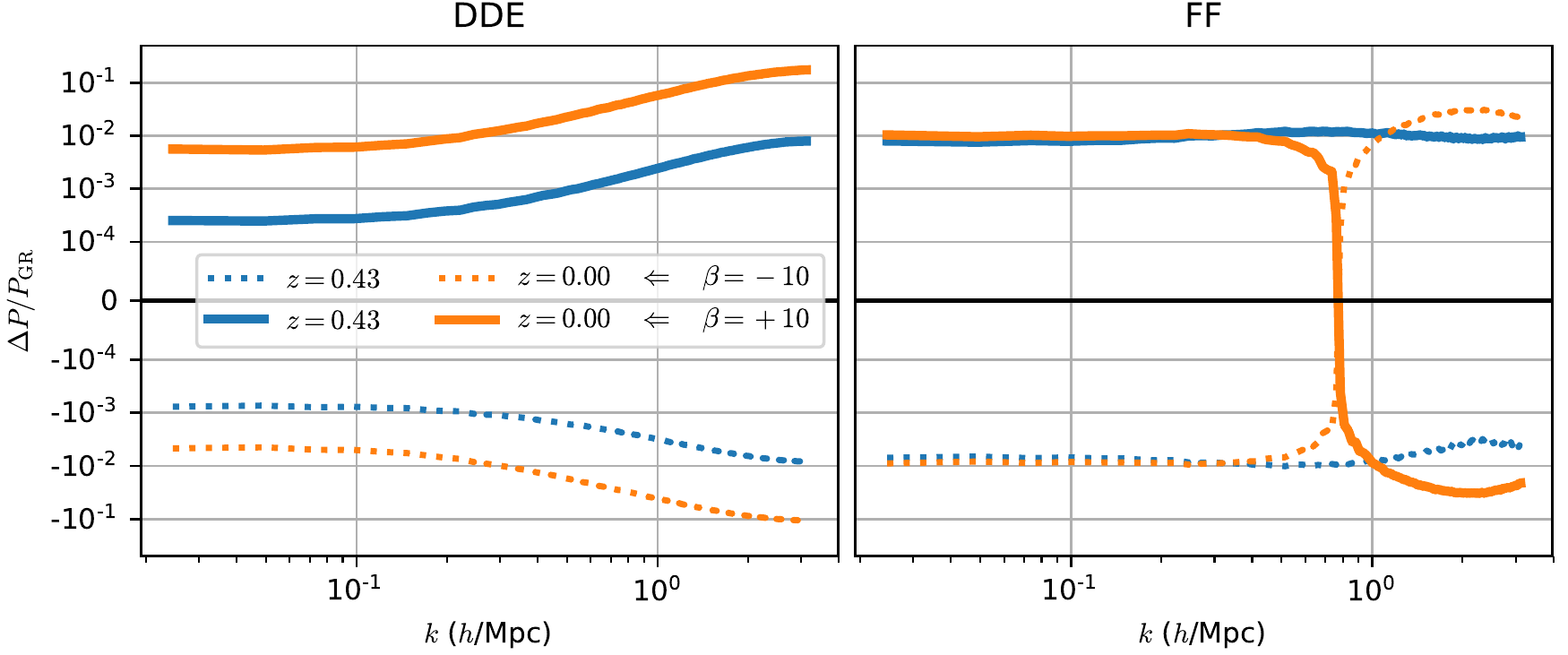}
  \caption{\label{fig:parameter_pk} Relative difference between dark matter power spectra of simulations with $\beta$ equal and different than zero.  The left and right panels correspond to the DDE and FF models respectively.  Different colours are different redshifts and different line styles correspond to simulations run with $\beta=-10$ and $\beta=10$.  The sharp transition between positive and negative values in the right panel is generated by the symlog scaling that we use to be able to plot positive and negative values.}
\end{figure*}

In this section, we discuss the impact that the parameter $\beta$, which determines the slope of the disformal coupling, has on the power spectrum of density perturbations.  Figure \ref{fig:parameter_pk} shows a comparison between simulations that were run with $\beta=0$ and $\beta=\pm 10$.  The left and right panels correspond to the base models DDE and FF respectively.  Continuous and dotted lines correspond to positive and negative values of $\beta$.

The left panel shows that allowing $\beta$ to be different than zero in the DDE model has a monotonic effect with frequency.  Positive values of $\beta$ force the coupling to go up with the value of the field (see Eq.~\ref{eq:B}), which in turn increases the power with respect to the base model defined by $\beta=0$.  As in this model, the power increases with respect to GR, the net effect is a faster evolution with respect to GR.  The opposite happens when $\beta$ is negative:  the coupling decreases as the field rolls down the potential, and thus reduces the impact of modified gravity in the power spectrum.  This decrease in power is relevant when it comes to use perturbations to further constrain the parameter space of this model.  In the previous section, we showed that even if the DDE model is compatible with CMB observations \citep{MifsudPlanck} it gives an increase in the power spectrum of about 10\% at $k\sim 3~h$/Mpc, which might be too large for it to be compatible with galaxy surveys.  This being a problem or not will naturally depend on the galaxy formation model used to connect these predictions with the actual distribution of galaxies.  In case this is indeed a problem, we show here that it can be alleviated by choosing negative values for $\beta$.  In fact, the relative difference between the simulation that was run with $\beta=-10$ and $\beta=0$ at the Nyquist frequency is of the same order than the effect that the base model with $\beta=0$ has with respect to GR and thus can reduce the modified gravity effects by a factor of about two.

The effects of $\beta$ in the FF model are more complex.  A positive coupling constant $\beta$ has the same effect in the power spectrum as in the DDE model at large scales (i.e. it induces an increase in power). However, at small scales, the impact of modified gravity in the power spectrum decreases with respect to the base model with $\beta=0$.  Note that the same happens when comparing the base model with GR, however the transition between positive and negative MG effects occurs at different frequency $k$.  The net effect of assuming positive $\beta$ is to increase even more the power with respect to GR at large scales and reduce it at small scales.  The opposite occurs when $\beta$ is assumed to be negative: there is a reduction of power at large scales and an increase at small scales.  As in the DDE case, allowing $\beta$ to move away from zero may help in reducing the problem that we found in previous section with excessive increase in power at large scales.  However, in this case, the effects of $\beta$ are one order of magnitude below the effect produced by the base model with $\beta=0$ and thus, including $\beta$ cannot save the model (at least not with the parameter we choose, which is equal to -10).

\section{Conclusions} \label{sec:conclusions}

We studied new phenomenology that arises when adding a disformal coupling to a very simple and well known extension to General Relativity (GR) such as the quintessence model.  The paper is divided in two main parts.  We first discuss analytical properties of the solutions of the field equation for the background as well as the perturbations.  In the second part, we present cosmological non-linear simulations that we run with the code \texttt{Isis} \citep{Isis}, which is based of the particle mesh code \texttt{Ramses} \citep{Ramses}.  We use the simulated data to show how our analytic predictions perform in realistic situations associated with the non-linear regime of cosmological evolution.

We start by describing background solutions of the Klein-Gordon equation for an Einstein-de Sitter cosmology.  We found that the shape of the solution depends on only two parameters $D$ and $F$ (see Eq.~\ref{defs_1} and \ref{defs_2}), which are combinations of the original four free parameters ($V_0$, $\nu$, $B_0$ and $\beta$).  In the case $F=0$, we identified three different characteristic time scales (shown in Table \ref{tab:definitions} and Figure \ref{fig:three_lines_background_solutions}), which determine the structure of the solution.  These scales consist in a transition between a disformal regime and a quintessence regime and a transition between these two regimes and a non-linear regime that occurs when the fields grow to the point that it is not possible to linearise the potential $V$.  Assuming $F\ne 0$ changes the definition of these time scales for $F>0$ and induces an instability in the case $F<0$, for which we provide a condition on its associated time scale.

The first part of the paper also deals with analytic properties of the perturbed Klein-Gordon equation.  In particular, we show that there is an approximate proportionality relation between the perturbed scalar field and the scalar perturbations of the metric.  The relation can be simplified by substituting the fields that constitute the proportionality factor with background quantities.  By analysing the modified geodesics equation, we also show that a similar relation exists between the Newtonian force and the fifth force that arises from the scalar field.  We close the analytic section by discussing the conditions under which the fifth force can be repulsive.  This is relevant in the context of the tension that is known to exist between measurements of the normalization of density perturbations in the universe using high and low redshift data sets \citep{PlanckLensing, PlanckDiscordance, 2017A&A...597A.126C, 2016MNRAS.459..971K}.

In the second part of the paper, we first used our N-body simulations to study properties of the field distribution.  Very good agreement was found between the exact solution obtained by the non-linear hyperbolic solver of the cosmological code (which can solve the equation without relying on any assumptions or approximations) and the prediction obtained following the proportionality relation with the metric perturbations.  This shows that, although the model is very complex, it can be simulated with a very simple algorithm based on the solution of the Poisson's equation, which exists in any standard gravity cosmological code.  We will present in a companion paper a detailed analysis of the accuracy with which the method can predict the evolution of matter.

An additional result associated with the simulated fields is that the perturbations do not grow monotonically with time, but can undergo oscillations.  Combining these results with our analytic results, we show that these oscillations cannot be sustained, but that field perturbations can have at most one maximum and one minimum and that ultimately will decrease towards zero.  The time scale for these oscillations naturally depends on the model parameters.

Our simulations also show that the repulsive forces that we predicted analytically do emerge in realistic situations.  We found that depending on the model parameters, repulsive forces can appear as transients that occur only at specific redshifts or be sustained in time.  In both cases, the existence of forces is related with the zeros of the function $\xi_0$, which connects metric and scalar field perturbations.

We complete the analysis of our simulations by studying how the fifth force that arises from the scalar field affects the matter distribution.  In particular, we focus on the power spectrum of density perturbations.  For the models that have $\beta=0$ (i.e. a constant disformal coupling), we show that models with weak coupling produce an increase of power at small scales, which is almost scale invariant at late times.  Models with larger coupling have an early increase in power which is followed with a decrease.  In these cases, the fifth force is such that it reduces the clustering with respect to GR.  We also studied the impact of $\beta$ in the density perturbations.  We found that positive or negative $\beta$, increase or decrease the effects of the fifth force at all scales.

Only one of all the simulated models (the DDE model) provides a background cosmology that is consistent with CMB data.  However, the aim of this work is not to provide best parameters, but to highlight new phenomenology.  The effects that we describe here may also appear in similar models which may be consistent with background data for different sets of parameters.  In other words, it may be possible to build models with effects similar to those discussed here and that at the same time provide a background that is consistent with data.


\section*{Acknowledgements}

Many thanks to Jack Morrice for helpful discussions and David Bacon and Kazuya Koyama for carefully reading the manuscript.  We also thank the Research Council of Norway for their support.  CLL acknowledges support from STFC consolidated grant ST/L00075X/1 \& ST/P000541/1 and ERC grant ERC-StG-716532-PUNCA.  The simulations used in this paper were performed on the NOTUR cluster FRAM. This paper is based upon work from COST action CA15117 (CANTATA), supported by COST (European Cooperation in Science and Technology).

\bibliographystyle{mnras}
\bibliography{references}

\appendix
\section{Asymptotic solutions for background fields for an Einstein-de Sitter universe}
\label{appendix:table}

\begin{table*}

  \aline  \vspace{-\bigskipamount} 
  \vspace{-\medskipamount}
  \begin{flalign}
    \intertext{\raggedright{\textbf{Q1. Quintessence, linear ($\pmb{\tilde{\chi}<<1}$), without damping}:}}
    \partial_{\tilde{t}}^2\tilde{\chi}  = 1
    \quad \quad
    & \Rightarrow 
    \quad \quad
    \tilde{\chi}  =  \frac{\tilde{t}^2}{2}
    \intertext{
      \vspace{-\bigskipamount}
      \begin{minipage}[h]{\textwidth}
        \begin{multicols}{2}
          The equation of motion is linear; its solution is a power law.
        \end{multicols}
        \vspace{-\bigskipamount}\aline  
      \end{minipage}
    }
    \intertext{\raggedright{\textbf{Q2. Quintessence, linear ($\pmb{\tilde{\chi}<<1}$), with damping}:}}
    \partial_{\tilde{t}}^2\tilde{\chi}  = -\frac{2}{\tilde{t}}\partial_{\tilde{t}}\tilde{\chi} + 1
    \quad \quad
    & \Rightarrow 
    \quad \quad
     \tilde{\chi} = \frac{\tilde{t}^2}{6}
    \intertext{
      \vspace{-\bigskipamount}
      \begin{minipage}[h]{\textwidth}
        \begin{multicols}{2}  
          The equation of motion is still linear.  The damping term is present at all times and changes only the normalization of the solution (not the slope).
        \end{multicols}
        \vspace{-\bigskipamount}\aline  
      \end{minipage}
    }
    \intertext{\raggedright{ \textbf{Q3. Quintessence, non-linear ($\pmb{\tilde{\chi}\sim 1}$), without damping}:}}
    \partial_{\tilde{t}}^2\tilde{\chi}  =  \exp\left(-\tilde{\chi}\right)
    \quad \quad
    & \Rightarrow 
    \quad \quad
    \tilde{\chi} =  \log\left[ \cosh^2\left(\frac{\tilde{t}}{\sqrt{2}} \right)  \right] 
    \sim \begin{cases}
      \tilde{t}^2/2 & \text{if $\tilde{t} \lesssim T^{\mathrm{nd}}_a$}, \\
      \sqrt{2} \tilde{t}  & \text{if $\tilde{t} \gtrsim T^{\mathrm{nd}}_a$}
    \end{cases}
    \intertext{
      \vspace{-\bigskipamount}
      \begin{minipage}[h]{\textwidth}
        \begin{multicols}{2}
          Taking into account the exponential definition of the potential gives a non-linearity to the equation which is active only for large enough values of $\tilde{\chi}$.  This non-linearity is negligible at early times and thus, the solutions behave as the linear solutions Q1 and Q2.  At late times, the force responsible for the evolution of the field decreases with time, giving a different slope to the solution.  The transition time $T^{\mathrm{nd}}_a$ between the linear and non-linear regimes can be 
          calculated as the moment in which the two asymptotes of the solution cross each other.
        \end{multicols}
        \vspace{-\bigskipamount}\aline  
      \end{minipage}
    }
    \intertext{\raggedright{ \textbf{Q4. Quintessence, non-linear ($\pmb{\tilde{\chi}\sim 1}$), with damping}:}}
    \partial_{\tilde{t}}^2\tilde{\chi}  = -\frac{2}{\tilde{t}}\partial_{\tilde{t}}\tilde{\chi} +  \exp\left(-\tilde{\chi}\right)
    \quad \quad
    & \Rightarrow
    \quad \quad
    \begin{aligned}
      \tilde{\chi} \sim
      \begin{cases}
        \tilde{t}^2/6 & \text{if $\tilde{t} \lesssim T_a$}, \\
        \log\left( \tilde{t}^2/2  \right)  & \text{if $\tilde{t} \gtrsim T_a$ (this is an exact solution)}
      \end{cases} 
    \end{aligned}
    \intertext{
      \vspace{-\bigskipamount}
      \begin{minipage}[h]{\textwidth}
        \begin{multicols}{2}
          A combination of the presence of the damping term and the reduction of the potential with time changes the slope of the solution in the non-linear regime, which becomes logarithmic at large times.  This changes slightly the transition time $T_a$, which we estimated numerically as the moment in which the logarithmic slope of the solution is the mean of the two asymptotes.  Note that the transition between linear and non-linear regimes and undamped and damped regimes occurs approximately at the same time. 
        \end{multicols}
        \vspace{-\bigskipamount}\aline  
      \end{minipage}
      \nonumber
    }
  \end{flalign}
  \caption{General properties of background solutions of the quintessence model.  See section \ref{sec:quint} for explanation.}
  \label{tab:quintessence}
\end{table*}


Tables \ref{tab:quintessence} and \ref{tab:disformal_beta_equal_zero} summarize solutions of the Klein-Gordon equation described in Sections \ref{sec:quint} and \ref{disformal_background_constant_coupling}.  These are solutions for the limits $(D,F)\rightarrow 0$ (Table \ref{tab:quintessence}) and $F\rightarrow 0$ (Table \ref{tab:disformal_beta_equal_zero}) for small and large values of the field and with and without including the damping term in the equation.

\begin{table*}

  \aline  \vspace{-\bigskipamount} 
  \vspace{-\medskipamount}  
  \begin{flalign}
      \intertext{\raggedright{\textbf{D1. Disformal, $\pmb{\beta=0}$, linear ($\pmb{\chi<<1}$), without damping}:}}
    \partial_{\tilde{t}}^2\tilde{\chi} = \frac{\tilde{t}^2}{\tilde{t}^2+D}
    \quad
    & \Rightarrow
    \quad
    \tilde{\chi} = \frac{\tilde{t}^2}{2} \left[ 1 - \frac{2\sqrt{D}}{\tilde{t}} \tan^{-1}\left(\frac{\tilde{t}}{\sqrt{D}}\right) + \frac{D}{\tilde{t}^2}\log \left(\frac{\tilde{t}^2 +D}{D}\right) \right] \sim 
    \begin{cases}
      \label{eq:early_disf}
      \tilde{t}^4/(12D) & \text{if $\tilde{t} \lesssim T^{\mathrm{nd}}_b$},\\
      \tilde{t}^2/2 & \text{if $\tilde{t} \gtrsim T^{\mathrm{nd}}_b$}
    \end{cases}
    \intertext{
      \vspace{-\bigskipamount}
      \begin{minipage}[h]{\textwidth}
        \begin{multicols}{2}
          The disformal coupling gives a time dependence to the term associated to the external force, which is such that it approaches zero at early times.  This changes the logarithmic slope of the solution with respect to the equivalent solution of the quintessence model (model Q1 in Table \ref{tab:quintessence}) at early times.  At late times, the time dependence approaches a constant and thus, the evolution of the field is as in the quintessence model Q1.  The transition time $T^{\mathrm{nd}}_b$ can be estimated as the moment where the two assymptotes cross each other. \\
        \end{multicols}
        \vspace{-\bigskipamount}\aline  
      \end{minipage}
    }
    \intertext{\raggedright{\textbf{D2. Disformal, $\pmb{\beta=0}$, linear ($\pmb{\chi<<1}$), with damping}:}}
    \partial_{\tilde{t}}^2\tilde{\chi} =  -2\frac{\tilde{t}}{\tilde{t}^2+D}\partial_{\tilde{t}}\tilde{\chi} + \frac{\tilde{t}^2}{\tilde{t}^2+D}
    \quad
    & \Rightarrow
    \quad
    \tilde{\chi} = \frac{\tilde{t}^2}{6} \left[ 1 - \frac{D}{\tilde{t}^2} \log \left( \frac{\tilde{t}^2 + D}{D}\right) \right] \sim 
    \begin{cases}
      \tilde{t}^4/(12D) & \text{if $\tilde{t} \lesssim T_b$},\\
      \tilde{t}^2/6 & \text{if $\tilde{t} \gtrsim T_b$}
      \label{disformal_no_beta_damping_small_field}
    \end{cases}
    \intertext{
      \vspace{-\bigskipamount}
      \begin{minipage}[h]{\textwidth}
        \begin{multicols}{2}  
          The damping term of the disformal equation differs with respect to the one we find in the quintessence case (model Q2 in Table \ref{tab:quintessence}) in that it approaches zero at early times.  Thus, the early time solution is not affected by the presence of this term (i.e. it is a power law with a logarithmic slope equal to four as in D1).  At late times, the solution switches to the solution for the quintessence model that includes damping (model Q2).  The transition occurs at a slightly earlier time $T_b$.      
        \end{multicols}
        \vspace{-\bigskipamount}\aline  
      \end{minipage}
    }
    \intertext{\raggedright{ \textbf{D3. Disformal, $\pmb{\beta=0}$, non-linear ($\pmb{\chi\sim 1}$), without damping}:}}
    \partial_{\tilde{t}}^2\tilde{\chi}  = \frac{\tilde{t}^2}{\tilde{t}^2+D}\exp\left(- \tilde{\chi}\right) 
    \quad
    & \Rightarrow
    \quad
    %
    \left\{
    \begin{aligned}
      \begin{array}{c}
        D\ll 1 \\
        \left(T^{\mathrm{nd}}_c \lesssim T^{\mathrm{nd}}_b\right)
      \end{array}
      \Rightarrow &
      ~~~~ \chi \sim
\begin{cases}
        \tilde{t}^4/(12D)           & \text{if $\tilde{t} \lesssim T^{\mathrm{nd}}_b$}\\
        \tilde{t}^2/2               & \text{if $T^{\mathrm{nd}}_b \lesssim \tilde{t} \lesssim T^{\mathrm{nd}}_a$} \\
        \sqrt{2} \tilde{t}  & \text{if $\tilde{t} \gtrsim T^{\mathrm{nd}}_a$} \\ 
      \end{cases} \\
      \begin{array}{c}
        D\gg 1 \\
        \left(T^{\mathrm{nd}}_c \gtrsim T^{\mathrm{nd}}_b\right)
      \end{array}
      \Rightarrow  &
      ~~~~ \chi \sim 3 \left[e^{-\tilde{t}^4/(12D)}-1\right] - \frac{3^{3/4}}{\sqrt{2}D^{1/4}}\Gamma\left[3/4,\frac{\tilde{t}^4}{12D},0 \right] \tilde{t} \sim 
      \begin{cases}
        \tilde{t}^4/(12D)                                   & \text{if $\tilde{t} \lesssim T^{\mathrm{nd}}_c$} \\
        \frac{3^{3/4} \Gamma(3/4)}{\sqrt{2}D^{1/4}}\tilde{t}  & \text{if $\tilde{t} \gtrsim T^{\mathrm{nd}}_c$}
      \end{cases}
  \end{aligned}
  \right.
  \intertext{
    \vspace{-\bigskipamount}
    \begin{minipage}[h]{\textwidth}
      \begin{multicols}{2}
For $D\ll 1$, the equation can be linearised with respect to $D$ and becomes a perturbed quintessence equation: 
$\partial_{\tilde{t}}^2\tilde{\chi} = \left(1-D/\tilde{t}^2\right)\exp\left(- \chi\right)$.  Since the disformal effects are a perturbation, the transition to the quintessence regime occurs before anything else happens.  Thus, the solution transitions first from the linear disformal to the linear quintessence regimes at $T^{\mathrm{nd}}_b$ (as in solution $D1$) and then towards the non-linear quintessence regime at $T^{\mathrm{nd}}_a$ (as in solution Q3). \\
For $D \gg 1$, the transition towards the non-linear regime occurs before the transition to the quintessence regime.  We can obtain a good approximation of the solution by linearizing the equation with respect to time and substituting the early time solution in the exponential function (i.e. by solving $\partial_{\tilde{t}}^2\tilde{\chi} = \tilde{t}^2\exp\left(- \tilde{t}^4/(12D)\right)$).  After the transition to the non-linear regime occured, the field behaves as a free particle and thus, no more transitions occur.
      \end{multicols}
      \vspace{-\bigskipamount}\aline  
    \end{minipage}
  }
     \intertext{\raggedright{ \textbf{D4. Disformal, $\pmb{\beta=0}$, non-linear ($\pmb{\chi\sim 1}$), with damping}:}}
    \partial_{\tilde{t}}^2\tilde{\chi}  = \frac{\tilde{t}^2}{\tilde{t}^2+D}\exp\left(- \tilde{\chi}\right) 
    \quad
    & \Rightarrow
    \quad
    %
    \left\{
    \begin{aligned}
      \begin{array}{c}
        D\ll 1 \\
        \left(T_c \lesssim T_b\right)
      \end{array}
      \Rightarrow &
      ~~~~ \chi \sim
      \begin{cases}
        \tilde{t}^4/(12D)           & \text{if $\tilde{t} \lesssim T_b$}\\
        \tilde{t}^2/6     & \text{if $T_b \lesssim \tilde{t} \lesssim T_a$} \\
        \log\left( \tilde{t}^2/2  \right)  & \text{if $\tilde{t} \gtrsim T_a$} \\ 
      \end{cases} \\
      \begin{array}{c}
        D\gg 1 \\
        \left(T_c \gtrsim T_b\right)
      \end{array}
      \Rightarrow  &
      ~~~~ \chi \sim
      \begin{cases}
        \tilde{t}^4/(12D)                                   & \text{if $\tilde{t} \lesssim T_c$} \\
        \frac{3^{3/4} \Gamma(3/4)}{\sqrt{2}D^{1/4}}\tilde{t}  & \text{if $T_c \lesssim \tilde{t} \lesssim T_b$} \\
        \log\left( \tilde{t}^2/2  \right) + 3^{5/4}\Gamma\left(3/4\right)D^{1/4}  & \text{if $ \tilde{t} \gtrsim T_b$} \\
      \end{cases}
  \end{aligned}
  \right.
  \intertext{
    \vspace{-\bigskipamount}
    \begin{minipage}[h]{\textwidth}
      \begin{multicols}{2}
For $D\ll 1$, the solution has a quick transition towards the linear quintessence regime as in D3.  Since damping was taken into account in these solutions, the transition is towards Q4 instead of Q3). \\
For $D \gg 1$, we have again (as in D3) that the transition towards the non-linear regime occurs before the transition to the quintessence regime.  The transition can be obtained by patching solutions of the different regimes, but taking into account that damping affects the quintessence regime at large $\tilde{t}$ and adding an appropiate constant to ensure continuity at $T_c$.
      \end{multicols}
      \vspace{-\bigskipamount}\aline  
    \end{minipage}
    \nonumber
}
  \end{flalign}
  \caption{General properties of background solutions of models with conformal coupling. \label{tab:disformal_beta_equal_zero}}
\end{table*}


\bsp    
\label{lastpage}

\end{document}